\documentclass[10pt,aps,prl,twocolumn,floatfix,showpacs,superscriptaddress]{revtex4-1}
\usepackage{graphicx,amsmath,amsfonts,mathrsfs}
\usepackage{bm}
\usepackage[]{color}

\begin{document}
\title{Microscopic  theory of electric polarization induced by skyrmionic order in GaV$_{4}$S$_{8}$}
\author{S.~A.~Nikolaev}
\email{saishi@inbox.ru}
\affiliation{National Institute for Materials Science, MANA,
1-1 Namiki, Tsukuba, Ibaraki 305-0044, Japan}
\author{I.~V.~Solovyev}
\email{SOLOVYEV.Igor@nims.go.jp}
\affiliation{National Institute for Materials Science, MANA,
1-1 Namiki, Tsukuba, Ibaraki 305-0044, Japan}
\affiliation{Department of Theoretical Physics and Applied Mathematics,
Ural Federal University, Mira St. 19, 620002 Yekaterinburg, Russia}
\date{\today}

\begin{abstract}
The lacunar spinel GaV$_{4}$S$_{8}$ was recently suggested to be a prototype multiferroic material hosting skyrmion lattice states with a sizeable polarization $\boldsymbol{P}$ coupled to magnetic order. We explain this phenomenon on the microscopic level. On the basis of density functional theory, we construct an effective model describing the behavior of magnetically active electrons in a weakly coupled lattice formed by \emph{molecular} orbitals of the (V$_{4}$S$_{4}$)$^{5+}$ clusters. By applying superexchange theory combined with the Berry-phase theory for $\boldsymbol{P}$, we derive a compass model relating the energy \emph{and polarization} change with the directions of spins $\boldsymbol{e}_{i}$ in magnetic bonds. We argue that, although each skyrmion layer is mainly formed by superexchange interactions in the same plane, the spin-dependence of $\boldsymbol{P}$ arises from the stacking misalignment of such planes in the perpendicular direction, which is inherent to the lacunar spinel structure. We predict a strong competition of isotropic, $\sim \boldsymbol{e}_{i}\boldsymbol{e}_{j}$, and antisymmetric, $\sim \boldsymbol{e}_{i} \times \boldsymbol{e}_{j}$, contributions to $\boldsymbol{P}$ that explains the experimentally observed effect.
\end{abstract}
\maketitle

\par \emph{Introduction}. In recent years, magnetic skyrmions~\cite{bog1,bog2}, topologically protected spin textures, have attracted high levels of interest due to their various potentials in the emerging field of spintronics~\cite{NagaosaTokura}. In most cases, they are stabilized by Dzyloshinskii-Moriya (DM) interactions in compounds with macroscopically broken inversion symmetry~\cite{sk1,sk5}. Owing to their topology and nanometer size, skyrmions behave as particle objects that can be moved over macroscopic distances by applying low-density electric currents~\cite{sk2,sk3} making them suitable candidates for applications in low-power nanoelectronics and data storage~\cite{sk4}.
\par Skyrmionic states have been theoretically predicted to occur in crystals belonging to certain crystallographic classes, which can be either polar or non-polar~\cite{bog1}. Being mostly observed in non-polar chiral structures, skyrmions in polar crystals are also of great interest due to their interplay with electric polarization, giving rise to fascinating multiferroic properties. Until recently, Cu$_{2}$OSeO$_{3}$ was the only known multiferroic material hosting a skyrmionic state~\cite{cuoseo1,cuoseo4}. Shortly after its first observation, an electric-field control of the skyrmion lattice in Cu$_{2}$OSeO$_{3}$ has been reported, indicating that many emergent properties of the skyrmion state can be tailored to the properties of a host material~\cite{cuoseo2,cuoseo3}. Overall, multiferroicity may give rise to many new prospects in a non-dissipative electric-field control of magnetic objects, and the existence of skyrmionic states in insulating magnetoelectric materials holds many potential applications for new-generation electronic devices.
\par Recently, a novel host material has been reported to exhibit these properties~\cite{gavs1}. GaV$_{4}$S$_{8}$ is a member of the lacunar spinel family with a non-centrosymmetric non-polar cubic $F\bar{4}3m$ structure, which at 38~K undergoes a structural transition to the polar rhombohedral $R3m$ phase~\cite{struc}, giving rise to the ferroelectric polarization $\sim 6000$ $\mu$C/m$^{2}$ along the rhombohedral direction $z\parallel[111]$~\cite{gavs2}. A complex phase diagram comprising paramagnetic, ferromagnetic (FM), skyrmion, and cycloidal states has been demonstrated, where the spin-driven excess polarization was assigned in each magnetic phase with a total value of $\sim 100$ $\mu$C/m$^{2}$, almost two orders of magnitude larger than that of Cu$_{2}$OSeO$_{3}$~\cite{gavs2}.
\par The existence of multiple ferroelectric phases in GaV$_{4}$S$_{8}$ indicates a complex interplay of charge, spin, and lattice degrees of freedom, making their theoretical description extremely important. Nevertheless, a rigorous theory of magnetoelectric coupling in skyrmion materials is lacking. It remains largely unknown what mechanisms are responsible for this coupling,  what aspects of the crystal structure play an essential role, and how a spin texture contributes to electric polarization in each ferroelectric phase. Thus, the purpose of this work is to fill this gap and explain the multiferroic properties of GaV$_{4}$S$_{8}$ on a microscopic level, through the rigorous Berry-phase theory of electric polarization combined with a realistic modeling approach.
\begin{figure}
\begin{center}
\includegraphics[width=0.49\textwidth]{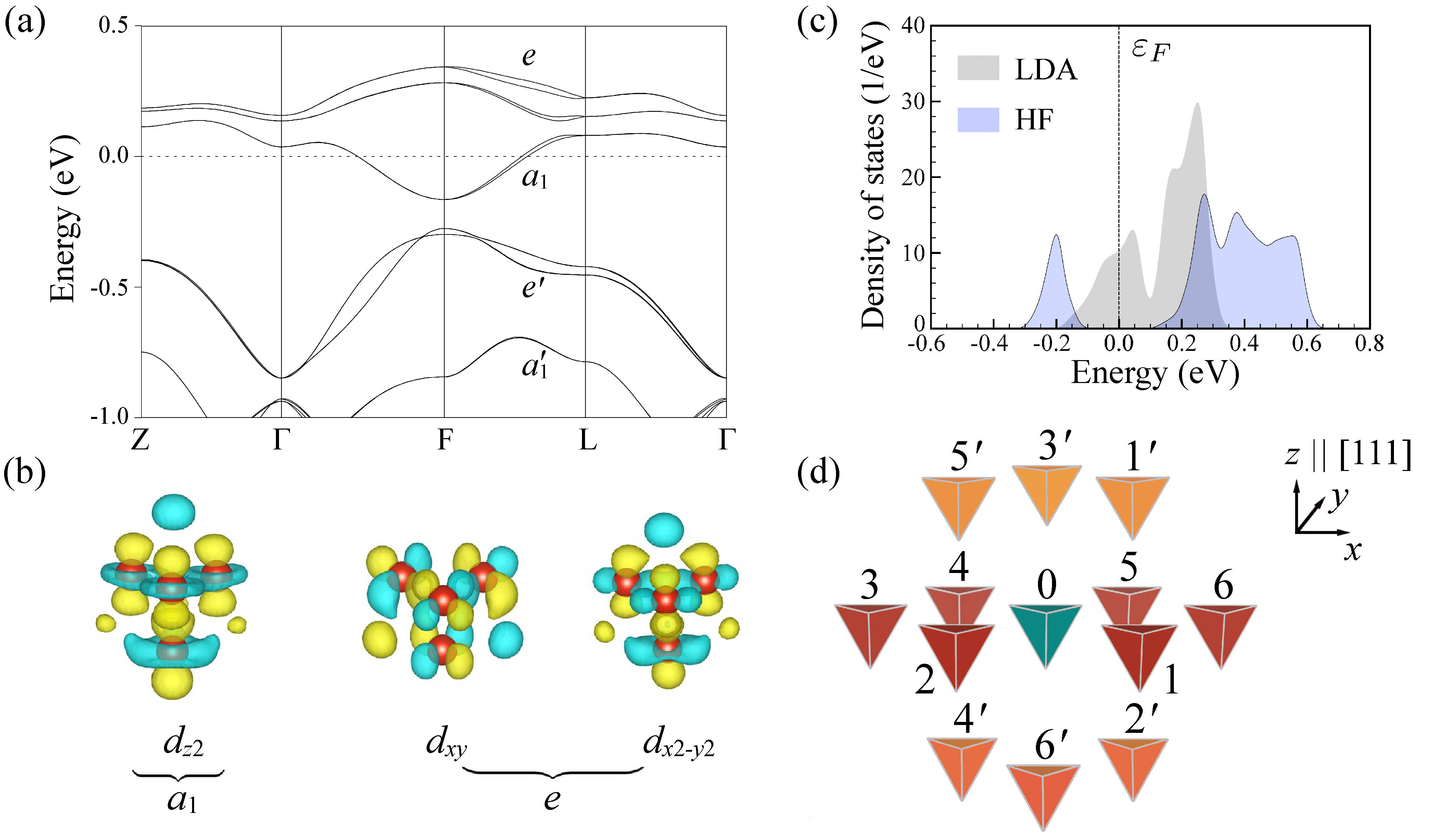}
\end{center}
\caption{(a)~Bands located near the Fermi level as calculated within LDA including spin-orbit coupling for the low-temperature GaV$_{4}$S$_{8}$. (b)~Wannier functions representing the high-lying $a_{1}$ and $e$ states. (c)~Density of states as obtained from LDA and Hartree-Fock calculations. (d)~Schematic view of the V$_{4}$ clusters with nearest neighbors.}
\label{fig:intro}
\end{figure}
\par \emph{Electronic model}. According to electronic structure calculations within local-density approximation (LDA)~\cite{lda}, as implemented in the VASP~\cite{vasp} and Quantum-ESPRESSO~\cite{qe} packages, the group of bands near the Fermi level is dominated by the V $3d$ states [Fig.~\ref{fig:intro}(a)], which strongly hybridize within each of the (V$_{4}$S$_{4}$)$^{5+}$ clusters, thus forming molecular-type orbitals. The hybridization between these molecular orbitals is considerably weak and leads to the weakly dispersive bands. In the $F\bar{4}3m$ phase, the molecular states belong to the $a'_{1}$, $e'$ and $t_{2}$ representations and are filled with seven electrons. Thus, the low-lying $a'_{1}$ and $e'_{\phantom{1}}$ states are double occupied and do not contribute to magnetism, while the highest 3-fold degenerate $t_{2}$ level accommodates one unpaired electron. The rhombohedral distortion in the $R3m$ phase lifts the degeneracy of the $t_{2}$ level, splitting it into a single $a_{1}$ and 2-fold degenerate $e$ states.
\par Carrying a local $S=\frac{1}{2}$ moment, the (V$_{4}$S$_{4}$)$^{5+}$ clusters can be regarded as magnetic building blocks, and the corresponding molecular $a_{1}\oplus e$ orbitals associated with the lattice of the V$_{4}$ tetrahedra can be chosen as a proper basis for the low-energy electronic model. In this regard, conventional band-structure methods may fail to properly include the electronic correlations between these composite molecular orbitals. Moreover, the complexity of skyrmion lattices, including hundreds of magnetic sites, is beyond the current abilities of \emph{ab-initio} techniques, and a model Hamiltonian approach turns out to be an essential tool to study the magnetic properties of GaV$_{4}$S$_{8}$. Thus, our first goal is to construct an effective Hubbard-type model while taking full advantage of the \emph{ab-initio} calculations in the Wannier basis,
\noindent
\begin{equation}
\hat{\mathcal{H}}^{\mathrm{el}}=\hat{\mathcal{H}}_{\mathrm{kin}}+\hat{\mathcal{H}}_{\mathrm{CF}}+\hat{\mathcal{H}}_{\mathrm{SO}}+\hat{\mathcal{H}}_{U}.
\label{eq:elmodel}
\end{equation}
\noindent The kinetic energy, $\hat{\mathcal{H}}_{\mathrm{kin}}=\sum_{i \ne j} \sum_{ab\sigma} t_{ij}^{ab}\hat{c}_{ia}^{\sigma\dagger}\hat{c}_{jb}^{\sigma\phantom{\dagger}}$, crystal-field splitting, $\hat{\mathcal{H}}_{\mathrm{CF}}=\sum_{i,a\in e,\sigma} \Delta\,\hat{c}_{ia}^{\sigma\dagger}\hat{c}_{ia}^{\sigma\phantom{\dagger}}$, and spin-orbit coupling (SOC) terms are identified through the matrix elements of the LDA Hamiltonian in the basis of \textit{molecular-type} Wannier orbitals~\cite{WannierRevModPhys,wan90}; $\hat{c}_{ia}^{\sigma\dagger}$ ($\hat{c}_{ia}^{\sigma\phantom{\dagger}}$) are the corresponding creation (annihilation) operators of an electron with spin $\sigma$ at site $i$ and orbital $a$ [$a=1$ stands for the $a_{1}=d_{z^{2}}$ orbital, and $a=2$ and $3$ stand for, respectively, $e = d_{xy}$ and $d_{x^{2}-y^{2}}$ orbitals, carrying also some weight of the $yz$ and $zx$ symmetry~\cite{Terakura}, as shown in Fig.~\ref{fig:intro}(b)].
\par The full set of model parameters is presented in Supplemental Material~\cite{SM}. The polar rhombohedral distortion gives rise to the crystal-field splitting, $\Delta = 98.1$ meV. The site-diagonal part of $\hat{\mathcal{H}}_{\mathrm{SO}}$ includes a conventional ``spherical'' term and the Rashba-type ($R$) contribution arising from the distortion~\cite{rashba}, $\hat{\mathcal{H}}_{\mathrm{SO}} = \zeta_{SO} \sum_{i} \hat{\boldsymbol{L}}_{i} \cdot \hat{\boldsymbol{S}}_{i} - \zeta_{SO}^{R} \sum_{i} \left( \hat{L}^{x}_{i}\hat{S}^{x}_{i} + \hat{L}^{y}_{i}\hat{S}^{y}_{i} \right)$, where the angular momentum operator is given in a compact form in terms of the antisymmetric Levi-Civita symbol as $( \hat{L}^{x}_{i} )^{ab} = -i\varepsilon_{2ab}$, $( \hat{L}^{y}_{i} )^{ab} = -i\varepsilon_{3ab}$, and $( \hat{L}^{z}_{i} )^{ab} = i\varepsilon_{1ab}$, and the calculated SOC constants are $\zeta_{SO} = 23.0$ meV and $\zeta_{SO}^{R} = 1.3$ meV. The theory of superexchange (SE) used below utilizes only those hopping parameters that involve the occupied $a_{1}$ orbital, $\vec{t}_{ij} = (t_{ij}^{11},t_{ij}^{12},t_{ij}^{13})$. For the in-plane bonds [$j$$=$$1$-$6$ in Fig.~\ref{fig:intro}(d)] these parameters are given by $\vec{t}_{0j} = (-1)^{j}t_{\parallel}^{S}(0, \sin \frac{\pi j}{3}, \cos \frac{\pi j}{3} ) + t_{\parallel}^{A}(\theta_{\parallel}, \cos \frac{\pi j}{3}, -$$\sin \frac{\pi j}{3} )$, where $t_{\parallel}^{S} = -$$25$ meV and $t_{\parallel}^{A} = -$$16$ meV stand for symmetric and antisymmetric parts, respectively, and $\theta_{\parallel} = 0.25$. For the out-of-plane bonds [$j$$=$$1'$-$6'$ in Fig.~\ref{fig:intro}(d)] we have $\vec{t}_{0j} = (-1)^{j}t_{\perp}^{S}(0, -$$\sin \frac{\pi j}{3}, \cos \frac{\pi j}{3} ) + t_{\perp}^{A}(\theta_{\perp}, -$$\sin \frac{\pi j}{3}, \cos \frac{\pi j}{3} )$, where $t_{\perp}^{S} = -$$23$ meV, $t_{\perp}^{A} = -$$22$ meV, and $\theta_{\perp} = 0.15$. Finally, the screened on-site Coulomb interactions
\begin{equation}
\hat{\mathcal{H}}_{U} = \frac{1}{2}
\sum_{i}  \sum_{\sigma \sigma'} \sum_{abcd} U^{abcd}
\hat{c}^{\sigma\dagger}_{i a } \hat{c}^{\sigma' \dagger}_{i c }
\hat{c}^{\sigma \phantom{\dagger}}_{i b }
\hat{c}^{ \sigma' \phantom{\dagger}}_{i d},
\label{eq:hublow}
\end{equation}
\noindent were evaluated by using the constrained random-phase approximation (cRPA)~\cite{rpa2}. The calculated values are $U\equiv U_{nnnn}=0.68$ eV and $J\equiv U_{nmmn}=0.08$ eV for the intraorbital Coulomb and Hund's rule exchange interactions, respectively. These values are not particularly strong because the molecular $t_{2}$ orbitals are rather extended in space, considerably reducing the bare interactions compared to their regular atomic values. Furthermore, the bare interactions are efficiently screened in cRPA due to the proximity of the target $t_{2}$ bands to the occupied $a'_{1}$ and $e'_{\phantom{1}}$ bands of the same V $3d$ character~\cite{JPCMreview}. Nevertheless, $U$ remains the largest parameter of the model that justifies the use of SE theory for constructing the spin model in the limit $\hat{t}_{ij}\ll U$~\cite{Anderson}.
\par The electronic model (\ref{eq:elmodel}) can be solved in the mean-field Hartree-Fock approximation~\cite{JPCMreview}, and the FM state with the indirect band gap of 0.15 eV is found to be the ground state for the low-temperature phase of GaV$_{4}$S$_{8}$ [Fig.~\ref{fig:intro}(c)]. Given the large hopping parameters between occupied $a_{1}$ and empty $e$ states, the FM ground state is also favoured by the Goodenough-Kanamori rule \cite{kanamori,goodenough}.
\par \emph{Spin model}. In the atomic limit, a single $t_{2}$ electron resides at the lowest Kramers doublet of $\hat{\mathcal{H}}_{\mathrm{CF}}+\hat{\mathcal{H}}_{\mathrm{SO}}$, $|\alpha_{i}\rangle$, and the corresponding Wannier function at site $i$, $| w_{i} \rangle = | \alpha_{i} \rangle$, specifies the direction of spin as $\boldsymbol{e}_{i} = \langle \alpha_{i} | \boldsymbol{\sigma} | \alpha_{i} \rangle / | \langle \alpha_{i} | \boldsymbol{\sigma} | \alpha_{i} \rangle |$. The inclusion of $\hat{t}_{ij}$ will induce the tails $|\alpha_{i \to j} \rangle$ of $| w_{i} \rangle$ spreading to neighboring sites $j$,
\begin{equation}
|w_{i} \rangle \approx |\alpha_{i} \rangle + |\alpha_{i \to j} \rangle,
\label{eq:tails}
\end{equation}
\noindent which can be evaluated within perturbation theory to 1st order in $\hat{t}_{ij}$ by considering virtual hoppings into the subspace of unoccupied states at neighbouring sites (and vice versa) as $|\alpha_{i \to j} \rangle = \hat{\cal M}_{j} \hat{t}_{ji} | \alpha_{i} \rangle$, where
$$
\hat{\cal M}_{j} = \sum\limits_{M} \frac{\hat{\cal P}_{j}| jM \rangle \langle jM | \hat{\cal P}_{j} }{E_{jM}},
$$
$E_{jM}$ and $| jM \rangle $ are, respectively, eigenvalues and eigenfunctions of the excited two-electron states at site $j$, constructed from $\hat{\mathcal{H}}_{\mathrm{CF}}+\hat{\mathcal{H}}_{\mathrm{SO}}+\hat{\mathcal{H}}_{U}$ in the basis of Slater determinants by using Slater-Condon rules, and $\hat{\cal P}_{j}$ is the projector operator in the form of two-electron Slater determinants, constructed from the occupied orbital $|\alpha_{j}\rangle$ and basis orbitals at site $j$ (thus enforcing the Pauli principle)~\cite{se1,se2,se3}. Then, the kinetic energy gain can be expressed as $E_{\mathrm{kin}} = \sum_{\langle ij \rangle} \left( \langle \alpha_{i} | \hat{t}_{ji} |\alpha_{i \to j} \rangle + i \leftrightarrow j \right)$. By considering all possible combinations of $| \alpha_{i} \rangle $ and $| \alpha_{j} \rangle $, corresponding to the $x$, $y$, and $z$ directions of spins at sites $i$ and $j$,  $E_{\mathrm{kin}}$ can be mapped onto the spin model $\mathcal{H}^{\mathrm{S}} = \sum_{\langle ij \rangle} \boldsymbol{e}_{i} \tensor{\mathscr{J}}_{ij} \boldsymbol{e}_{j}$, which is further rearranged as~\cite{SM}
\begin{equation}
\mathcal{H}^{\mathrm{S}} = \sum\limits_{\langle i j\rangle} \left( - J_{ij}\boldsymbol{e}_{i}\boldsymbol{e}_{j} + \boldsymbol{D}_{ij} \boldsymbol{e}_{i}\times\boldsymbol{e}_{j} + \boldsymbol{e}_{i} \tensor{\Gamma}_{ij} \boldsymbol{e}_{j} \right),
\label{eq:spinmodel}
\end{equation}
\noindent in terms of the isotropic exchange constants $J_{ij}$, antisymmetric DM vectors $\boldsymbol{D}_{ij}$, and the traceless symmetric anisotropic tensors $\tensor{\Gamma}_{ij}$. Using parameters of the electronic model~(\ref{eq:elmodel}), we obtain: $J_{\parallel} = $ $0.180$ meV and $J_{\perp} = 0.217$ meV for the nearest-neighbor in-plane and out-of-plane interactions, respectively [$j$$=$$1$-$6$ and $1'$-$6'$ in Fig.~\ref{fig:intro}(d)]. The corresponding Curie temperature $T_{\rm C}\sim10$~K estimated in random phase approximation~\cite{tyab} is close to the experimental value of 13~K. The resulting DM interactions can be written in a compact form as $\boldsymbol{D}_{0j} = d_{\parallel} \left( \sin \frac{\pi j}{3}, \cos \frac{\pi j}{3}, (-1)^{j} \delta \right)$ for $j$$=$$1$-$6$, where $d_{\parallel} = 0.073$ meV and $\delta = 0.137$, and $\boldsymbol{D}_{0j} = d_{\perp} ( \cos \frac{\pi j}{3}, \sin \frac{\pi j}{3}, 0 )$ for $j$$=$$1'$-$6'$, where $d_{\perp} = 0.057$ meV. The parameters of $\tensor{\Gamma}_{ij}$ can be neglected on account of their smallness~\cite{SM}.
\par \emph{Electric polarization}. The theory of SE interactions is well established and constitutes the basis of the so-called anisotropic compass model, which is widely used for the analysis of magnetic properties of $5d$ iridium oxides~\cite{Khaliullin2009}. In the following, we formulate a similar anisotropic compass model for electric polarization. The rigorous Berry-phase theory relates the polarization change with expectation values of the position operator, calculated in the Wannier functions basis for the occupied states~\cite{ber2}
\begin{equation}
\boldsymbol{P}=-\frac{e}{V}\sum\limits_{i}^{\mathrm{occ}} \langle w_{i} | \boldsymbol{r} | w_{i} \rangle,
\label{eq:elpol}
\end{equation}
\noindent where $-$$e$ and $V$ is the electron charge and the unit cell volume, respectively. By this definition, all spin dependencies of $\boldsymbol{P}$ are incorporated in $| w_{i} \rangle$, so one needs to evaluate the change in the distribution of $| w_{i} \rangle$ caused by the change of magnetic order. In the lattice model, this change can be described by the tails of Wannier functions, $|\alpha_{i \to j} \rangle$, spreading to neighboring sites. Then, substituting Eq.~(\ref{eq:tails}) in Eq.~(\ref{eq:elpol}), electric polarization can be expressed as a sum of bond contributions $\boldsymbol{P}=\sum_{\langle ij \rangle} \boldsymbol{P}_{ij}$~\cite{footnote1}, where
\begin{equation}
\boldsymbol{P}_{ij} = \frac{e}{V} \boldsymbol{\tau}_{ji} \left( \langle \alpha_{j \to i} | \alpha_{j \to i} \rangle - \langle \alpha_{i \to j} | \alpha_{i \to j} \rangle \right),
\end{equation}
\noindent and $\boldsymbol{\tau}_{ji}=\boldsymbol{R}_{j}-\boldsymbol{R}_{i}$ is the bond vector connecting neighbouring sites~\cite{pol1,pol2,superpol}. The quantity $\langle \alpha_{i \to j} | \alpha_{i \to j} \rangle$, which is nothing else but the Wannier weight transfer from site $i$ to site $j$, can be obtained in the framework of SE theory as $\langle \alpha_{i \to j} | \alpha_{i \to j} \rangle = \langle \alpha_{i} | \hat{t}_{ij}\hat{\cal M}^{2}_{j} \hat{t}_{ji} | \alpha_{i} \rangle$. By considering different directions of spins for $| \alpha_{i} \rangle $ and $| \alpha_{j} \rangle $, the spin-driven part of electric polarization can be written as $\boldsymbol{P} = \sum_{\langle ij \rangle} \boldsymbol{\epsilon}_{ji} \left( \boldsymbol{e}_{i}\tensor{\mathscr{P}}_{ij}\boldsymbol{e}_{j} \right)$ or
\begin{equation}
\boldsymbol{P} = \sum\limits_{\langle i j\rangle} \boldsymbol{\epsilon}_{ji} \left( P_{ij}\boldsymbol{e}_{i}\boldsymbol{e}_{j} + \boldsymbol{\mathcal{P}}_{ij} \boldsymbol{e}_{i}\times\boldsymbol{e}_{j} + \boldsymbol{e}_{i} \tensor{\Pi}_{ij} \boldsymbol{e}_{j} \right),
\label{eq:spinpol}
\end{equation}
\noindent where $\boldsymbol{\epsilon}_{ji} = \boldsymbol{\tau}_{ji}/|\boldsymbol{\tau}_{ji}|$. This is an analogue of Eq.~(\ref{eq:spinmodel}), where $\boldsymbol{\epsilon}_{ji} P_{ij}$, $\boldsymbol{\epsilon}_{ji} \boldsymbol{\mathcal{P}}_{ij}$, and  $\boldsymbol{\epsilon}_{ji} \tensor{\Pi}_{ij}$ stand for isotropic, antisymmetric, and anisotropic symmetric contributions, respectively~\cite{footnote2}. Importantly, since $\boldsymbol{P}_{ij} \parallel \boldsymbol{\epsilon}_{ji}$, \emph{only the out-of-plane bonds can contribute to the polarization change along} $z$.
\par In order to clarify the microscopic origin of electric polarization in GaV$_{4}$S$_{8}$, it is useful to consider an analytical expression for $P_{ij}$, which can be easily obtained in the absence of SOC. To 1st order in $J/(U+\Delta)$, it yields~\cite{SM}: $P_{ij} \approx  (e |\boldsymbol{\tau}_{ji}|/V) \mathcal{T}_{ij} J/(U+\Delta)^3$, where $\mathcal{T}_{ij} = (t_{ji}^{12})^{2}+(t_{ji}^{13})^{2}-(t_{ij}^{12})^{2}-(t_{ij}^{13})^{2}$ is the antisymmetric tensor ($\mathcal{T}_{ij} =-\mathcal{T}_{ji}$). Thus, in order to have finite $P_{ij}$, it is essential that (i) the Hund's rule coupling $J$ should be finite; and (ii) inversion symmetry of the bond connecting neighbouring sites $i$ and $j$ should be crystallographically broken (otherwise, $\mathcal{T}_{ij} = \mathcal{T}_{ji}$ and, therefore, $\mathcal{T}_{ij} = 0$, as indeed happens in the high-temperature $F\bar{4}3m$ phase). These two properties hold even in the presence of SOC. Particularly, if $J=0$, the entire tensor $\tensor{\mathscr{P}}_{ij}$ is identically equal to zero, as confirmed by our calculations. Furthermore, for equivalent bonds in the positive ($+$) and negative ($-$) directions of $z$, we have $\mathcal{T}^{-} = - \mathcal{T}^{+}$, which is the direct consequence of translational invariance and the antisymmetry of $\mathcal{T}_{ij}$. In combination with $\boldsymbol{\epsilon}_{ji} = -\boldsymbol{\epsilon}_{ij}$, it results in a finite contribution to $\boldsymbol{P}$.
\par The calculated parameters for $j$$=$$1'$-$6'$ are $P_{0j} = (-1)^{j}P_{\perp}$ and $\boldsymbol{\mathcal{P}}_{0j} = (-1)^{j} p_{\perp}(\cos \frac{\pi j}{3}, \sin \frac{\pi j}{3}, 0)$, where $P_{\perp} = -$$362$ $\mu\mathrm{C/m}^{2}$ and $p_{\perp}= 41$ $\mu\mathrm{C/m}^{2}$. As we will see below, they are mainly responsible for the magnetic state dependence of $P^{z}$. The corresponding polarization in the FM phase is calculated from Eq.~(\ref{eq:spinpol}) as $P^{z} = 3 \epsilon_{01'}^{z}P_{\perp} = 889$ $\mu\mathrm{C/m}^{2}$ (where $\epsilon_{01'}^{z} = 0.819$~\cite{struc}), while its thermal average in the paramagnetic state yields $P^{z}=0$. As a result, we expect a large spin-driven excess polarization in the FM phase. The effect is very generic and can readily take place in other polar magnets~\cite{Fe2Mo3O8.1,Fe2Mo3O8.2}. For the in-plane bonds $j$$=$$1$-$6$, we have $P_{\parallel} \equiv 0$ and $\boldsymbol{\mathcal{P}}_{0j} = (-1)^{j} p_{\parallel}(\cos \frac{\pi j}{3}, -\sin \frac{\pi j}{3}, 0)$, where $p_{\parallel}= 30$ $\mu\mathrm{C/m}^{2}$. Since $\epsilon_{0j}^{z}=0$, these bonds do not contribute to $P^{z}$. Nevertheless, $\boldsymbol{\mathcal{P}}_{0j}$ can give rise to small $P^{x,y}$, provided that the symmetry is lowered by magnetic order, as in the proper-screw spin spiral~\cite{superpol,MnI2_Kurumaji}. Finally, in the multidomain samples~\cite{gavs1} the value of $P^{z}$ will be deteriorated: in the domains $[1\bar{1}\bar{1}]$, $[\bar{1}1\bar{1}]$, and $[\bar{1}\bar{1}1]$, $\boldsymbol{P}$ is parallel to the corresponding rhombohedral axes, whose $z$ component is opposite to the one of the main domain $[111]$. Moreover, since $P_{\parallel} = 0$, there are no other contributions to $P^{z}$ coming from the domains $[1\bar{1}\bar{1}]$, $[\bar{1}1\bar{1}]$, and $[\bar{1}\bar{1}1]$. This can explain a relatively small value of spin-driven polarization ($\sim$$100$ $\mu\mathrm{C/m}^{2}$) observed experimentally~\cite{gavs2}, in comparison with the results of our theoretical calculations.
\par
\begin{figure}[t!]
\begin{center}
\includegraphics[width=0.49\textwidth]{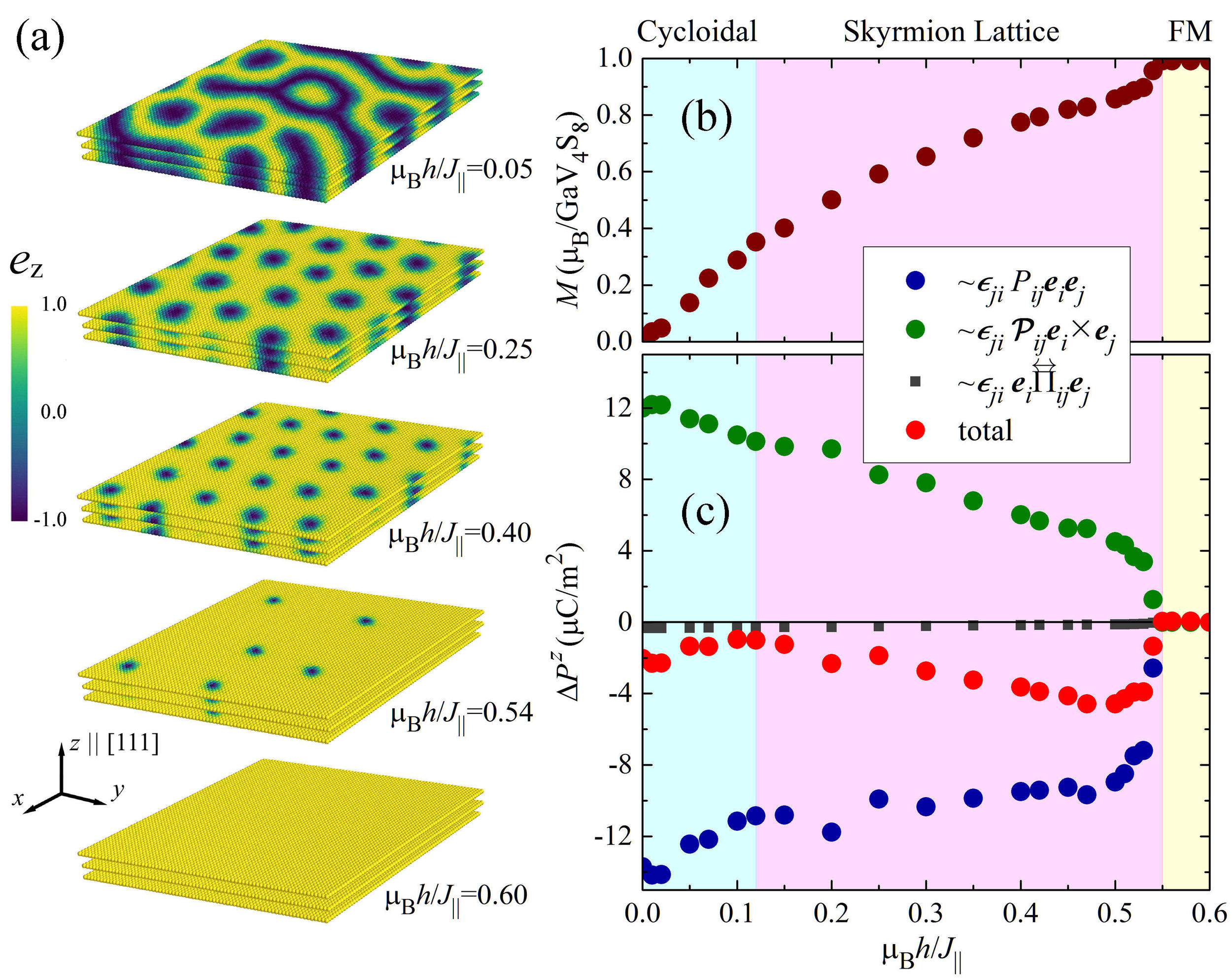}
\end{center}
\caption{(a)~Spin patterns as obtained in Monte Carlo calculations for the model (\ref{eq:spinmodel}) with an applied magnetic field $h \parallel z$ at temperature $T=0.1J_{\parallel}$ for the supercell of $72$$\times$$72$$\times$$3$ spins with periodic boundary conditions. In these notations,  a ``skyrmion lattice'' means the lattice of well distinguished skyrmionic tubes of the same size, while a ``cycloidal phase'' includes large interconnected regions with the same direction of spins. The corresponding $h$-dependence of (b)~the magnetization and (c)~electric polarization: total and partial contributions, calculated from Eq.~(\ref{eq:spinpol}) relative to the FM state.}
\label{fig:pol}
\end{figure}

 \emph{Phase diagram}. We perform classical Monte Carlo calculations for the spin model~(\ref{eq:spinmodel}) with an applied magnetic field $h\parallel z$ by using a heat-bath algorithm combined with overrelaxation~\cite{SM,montecarlo}. In these calculations, we assume that DM interactions are mainly responsible for the in-plane non-collinear alignment of spins and neglect possible spatial modulations of the magnetic textures along $z$. This is consistent with experimental neutron scattering data~\cite{gavs1} that report no magnetic superstructures along the direction of the magnetic field. The results calculated for supercells with the minimal periodicity along $z$ are shown in Fig.~\ref{fig:pol}(a) and nicely reproduce the main sequence of cycloidal $\to$ skyrmionic $\to$ FM states in the phase diagram of GaV$_{4}$S$_{8}$ with the increase of $h$~\cite{gavs2}. As seen, the two-dimensional spin patterns tend to stack ferromagnetically along $z$, that is naturally explained by $J_{\perp}$. In the GaV$_{4}$S$_{8}$ structure, this stacking of monolayers is misaligned by the rhombohedral translations, so that the adjacent skyrmionic layers experience an additional shift in the $xy$ plane. Therefore, there will always be some noncollinearity of spins between the adjacent layers~\cite{stacking} that, according to Eq.~(\ref{eq:spinpol}), will contribute to the excess spin-driven polarization.  
\par In order to describe this effect quantitatively, we evaluate the total and partial contributions to $P^{z}$ by using Eq.~(\ref{eq:spinpol}) and the distribution of spins $\{ \boldsymbol{e}_{i} \}$ obtained in Monte Carlo calculations. The results are summarized in Fig.~\ref{fig:pol}(c), where we use the FM state as the reference point. Particularly, we note a strong competition of the isotropic ($\sim$$\boldsymbol{e}_{i} \boldsymbol{e}_{j}$) and antisymmetric ($\sim$$\boldsymbol{e}_{i}\times\boldsymbol{e}_{j}$) contributions, while the anisotropic symmetric part ($\sim$$\boldsymbol{e}_{i} \tensor{\Pi}_{ij} \boldsymbol{e}_{j}$) is negligibly small. As expected, the antisymmetric contribution decreases with the increase of $h$ and vanishes in the collinear FM state. On the contrary, the isotropic contribution takes its maximal value in the FM state and is further reduced by a noncollinear alignment of spins. Since the change of $\boldsymbol{e}_{i}\times\boldsymbol{e}_{j}$ and $\boldsymbol{e}_{i} \boldsymbol{e}_{j}$ is proportional to $\phi_{ij}$ and $\phi_{ij}^2$, respectively (with $\phi_{ij}$ being the angle between $\boldsymbol{e}_{i}$ and $\boldsymbol{e}_{j}$, which is induced by DM interactions and proportional to $\zeta_{SO}$), both the isotropic and antisymmetric mechanisms are of 2nd order in $\zeta_{SO}$, while the change of $\boldsymbol{e}_{i} \tensor{\Pi}_{ij} \boldsymbol{e}_{j}$ is only of 3rd order. This naturally explains the hierarchy of partial contributions to $\Delta P^z$ in Fig.~\ref{fig:pol}(c). Furthermore, the antisymmetric mechanism dominates when the skyrmions are large and the spin texture slowly varies in space. In this region, electric polarization decreases with $h$, in agreement with the experimental observation~\cite{gavs2}. The corresponding polarization change of about $4$ $\mu\mathrm{C/m}^{2}$ is also consistent with experimental data~\cite{gavs2}. Finally, our conclusion clearly differs from the phenomenological analysis presented in~\cite{gavs2}, arguing that the antisymmetric DM interactions are solely needed to stabilize the cycloidal and skyrmion phases, while the corresponding polarization change is driven by the isotropic and anisotropic symmetric terms. In fact, we also expect a small region in the phase diagram, where the magnetization is nearly saturated [Fig.~\ref{fig:pol}(b)] and the skyrmion size is small, so the polarization change is mainly governed by the isotropic mechanism and is expected to increase with $h$. Overall, the $h$ dependence of spin-driven polarization in the skyrmion phase depends on the skyrmion size and the way a skyrmion lattice is packed, leading to different competing scenarios. Finally, it is worth noting that spin structure modulations driven by the out-of-plane DM interactions may also take place and increase the antisymmetric contribution to spin-driven polarization.
\par \emph{Conclusion.} We have presented the microscopic theory of spin-driven electric polarization in GaV$_{4}$S$_{8}$. Based on the realistic model derived from first-principles electronic structure calculations, we have shown that the spin-excess polarization along the rhombohedral $z$ axis associated with the ferromagnetic, skyrmionic, and cycloidal states, is given by the \emph{interlayer} electron transfer and originates from the strongly competing isotropic and antisymmetric contributions. The proposed theory is very general and can be applied to other multiferroic materials, including those hosting skyrmionic states.


\begin{thebibliography}{99}
\bibitem{footnote1}
The on-site contribution $\langle \alpha_{i} | \boldsymbol{r} | \alpha_{i} \rangle$ to the polarization change was found to be small~\cite{SM}.
\bibitem{stacking}
For example, if the skyrmion center is translated along the bond $0$-$1'$ [Fig.~\ref{fig:intro}(d)], the neighbouring out-of-plane spins will stay non-collinear in two other bonds $0$-$3'$ and $0$-$5'$. These interlayer translations may take place randomly.
\end{thebibliography}
\end{document}


\title{Supplemental Materials:\\ Microscopic theory of electric polarization induced by skyrmionic order in GaV$_{4}$S$_{8}$}
\author{S.~A.~Nikolaev}
\email{saishi@inbox.ru}
\affiliation{International Center for Materials Nanoarchitectonics, National Institute for Materials Science,
1-1 Namiki, Tsukuba, Ibaraki 305-0044, Japan}
\author{I.~V.~Solovyev}
\email{SOLOVYEV.Igor@nims.go.jp}
\affiliation{International Center for Materials Nanoarchitectonics, National Institute for Materials Science,
1-1 Namiki, Tsukuba, Ibaraki 305-0044, Japan}
\affiliation{Department of Theoretical Physics and Applied Mathematics,
Ural Federal University, Mira St. 19, 620002 Yekaterinburg, Russia}
\maketitle

\begin{figure*}
\begin{center}
\includegraphics[width=0.80\textwidth]{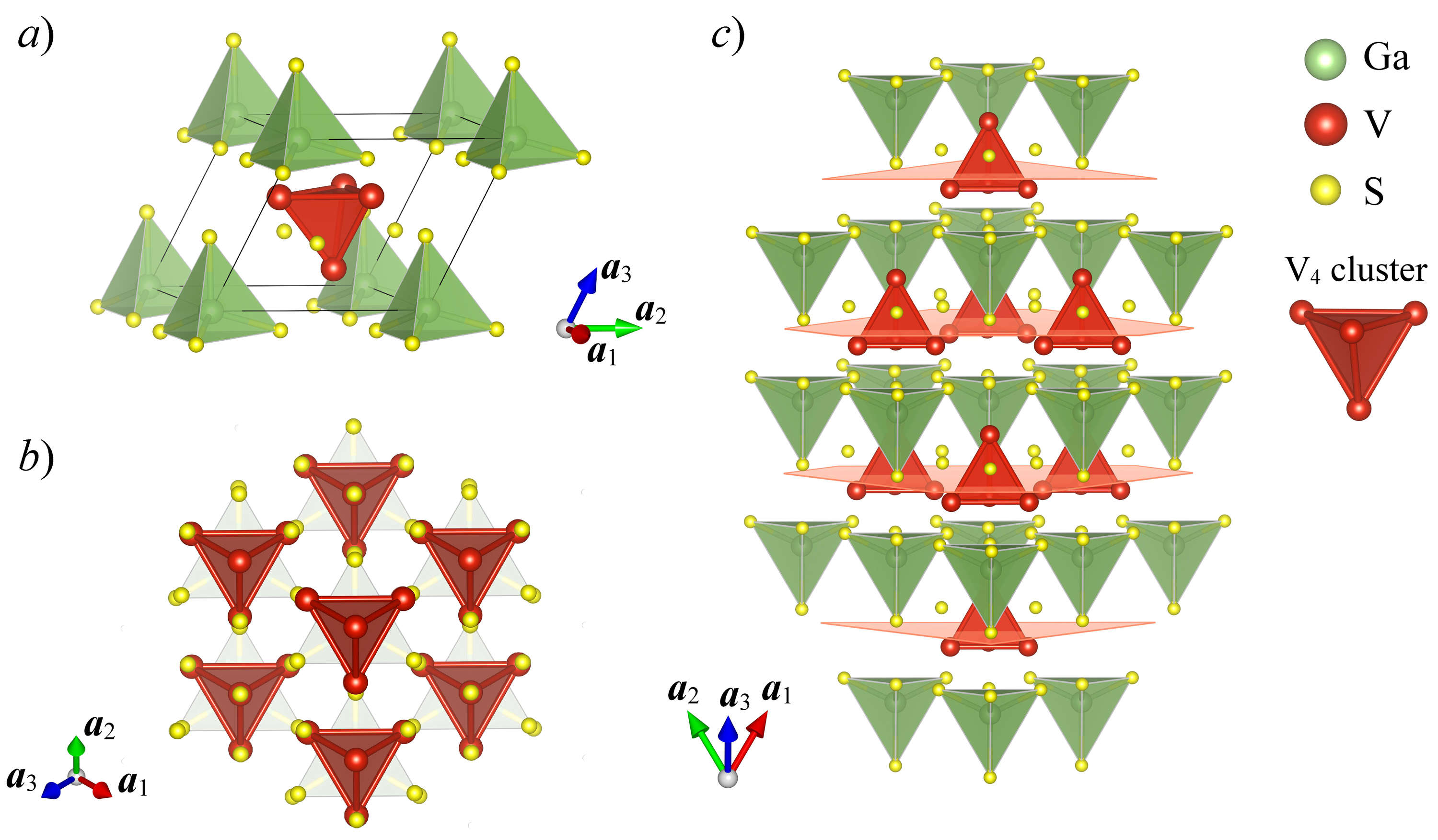}
\end{center}
\caption{Crystal structure of GaV$_{4}$S$_{8}$ visualised by VESTA software \cite{vesta}: $a$) unit cell, and $b,c$) planes along the [111] direction.}
\label{fig:struc}
\end{figure*}

\section{First-principles calculations}

\par In our studies, we adopt the experimental crystallographic data of GaV$_{4}$S$_{8}$ shown in Table~I and Fig.~\ref{fig:struc}. Electronic structure calculations were performed within the conventional local-density approximation (LDA) \cite{lda} and projected augmented wave \cite{paw} formalisms, as implemented in the Vienna ab-initio simulation package (VASP)~\cite{vasp}. Band structure calculations including spin-orbit coupling (LDA+SO) were carried out by using norm-conserving pseudopotentials as implemented in the Quantum- ESPRESSO package (QE)~\cite{qe}. The plane wave cutoff was set to 1360 eV, the Brillouin zone was sampled by a $16\times16\times16$ Monkhorst-Pack $k$-point mesh \cite{mp}, and the convergence criteria for the total energy calculations was 10$^{-9}$ eV. To calculate hopping parameters, we employed the maximally localized Wannier functions as implemented in the Wannier90 package~\cite{wanapp,wan90}. The resulting electronic structures and model parameters calculated in VASP and QE without spin-orbit coupling are in good agreement.

\par The on-site Coulomb parameters were calculated within the constrained random-phase approximation defined in the Wannier functions basis \cite{rpa1,rpa2}:

\begin{equation}
U_{i}^{\alpha\beta\gamma\delta}=\iint d\boldsymbol{r}d\boldsymbol{r}'w_{i}^{\dagger\alpha}(\boldsymbol{r})w_{i}^{\beta}(\boldsymbol{r})U(\boldsymbol{r},\boldsymbol{r}')w_{i}^{\dagger\gamma}(\boldsymbol{r}')w_{i}^{\delta}(\boldsymbol{r}')=\langle w_{i}^{\alpha} w_{i}^{\gamma}|U| w_{i}^{\beta} w_{i}^{\delta} \rangle,
\end{equation}

\noindent where the Greek superscript takes on the number of orbitals. Here, $U(\boldsymbol{r},\boldsymbol{r}')$ is the partially screened Coulomb interaction for the target bands comprising the low-energy model:

\begin{equation}
U = W(1+P_{d}W)^{-1}
\end{equation}

\noindent and

\begin{equation}
W = (1-vP)^{-1}v,
\end{equation}

\noindent where $v$ and $W$ are the bare and fully screened Coulomb interactions, respectively; $P$ is the total polarization including transitions between all states $\{\psi_{i},\varepsilon_{i}\}$:

\begin{equation}
P(\boldsymbol{r},\boldsymbol{r}')=2\sum\limits_{n}^{occ}\sum\limits_{n'}^{unocc}\psi_{n}^{*}(\boldsymbol{r})\psi_{n'}^{}(\boldsymbol{r})\psi_{n'}^{*}(\boldsymbol{r'})\psi_{n}^{}(\boldsymbol{r'})\left\{\frac{1}{\varepsilon_{n}-\varepsilon_{n'}+i\delta} + \frac{1}{\varepsilon_{n}-\varepsilon_{n'}-i\delta} \right\}
\end{equation}

\noindent that can be further divided into $P=P_{r}+P_{d}$, so that $P_{d}$ includes all possible transitions between the target bands, and $P_{r}$ is the rest of the polarization. This method was implemented in VASP.

\begin{table*}[t]
\caption{Crystal structure parameters and Wyckoff positions of GaV$_{4}$S$_{8}$, as obtained from X-ray powder data \cite{struc}.}
\begin{center}
\begin{tabular}{c|lcc}
\hline
\hline
\multicolumn{4}{c}{High-temperature phase $F\bar{4}3m$, $a=9.661$ \AA}\\
\hline
Ga &\qquad \,4a \,(0,0,0) &  \multicolumn{2}{c}{-} \\
V   & \qquad 16e (x,x,x) & \multicolumn{2}{c}{x = 0.6060}\\
S1 & \qquad 16e (x,x,x) & \multicolumn{2}{c}{x = 0.3707}\\
S2 & \qquad 16e (x,x,x) & \multicolumn{2}{c}{x = 0.8642}\\
\hline
\multicolumn{4}{c}{Low-temperature phase $R3m$, $a=6.799$ \AA, $c=16.782$ \AA}\\
\multicolumn{4}{c}{($a_{\mathrm{rh}}=6.834$ \AA, $\alpha_{\mathrm{rh}}=59.66^{\circ}$)} \\
\hline
Ga & \qquad 3a \, (0,0,z) &\qquad - & z = 0.0\\
V1 & \qquad 3a \, (0,0,z) &\qquad - & z = 0.3903\\
V2 & \qquad 9b (x,2x,z) &\qquad x = 0.1943 & z = 0.2001\\
S1 & \qquad 3a \, (0,0,z) &\qquad - & z = 0.6285\\
S2 & \qquad 9b (x,2x,z) &\qquad x = 0.1686& z = 0.4559\\
S3 & \qquad 9b (x,2x,z) &\qquad x = 0.1827& z = 0.9517\\
S4 & \qquad 3a \, (0,0,z) &\qquad - & z = 0.1384\\
\hline
\hline
\end{tabular}
\end{center}
\end{table*}

\begin{figure*}
\begin{center}
\includegraphics[width=0.85\textwidth]{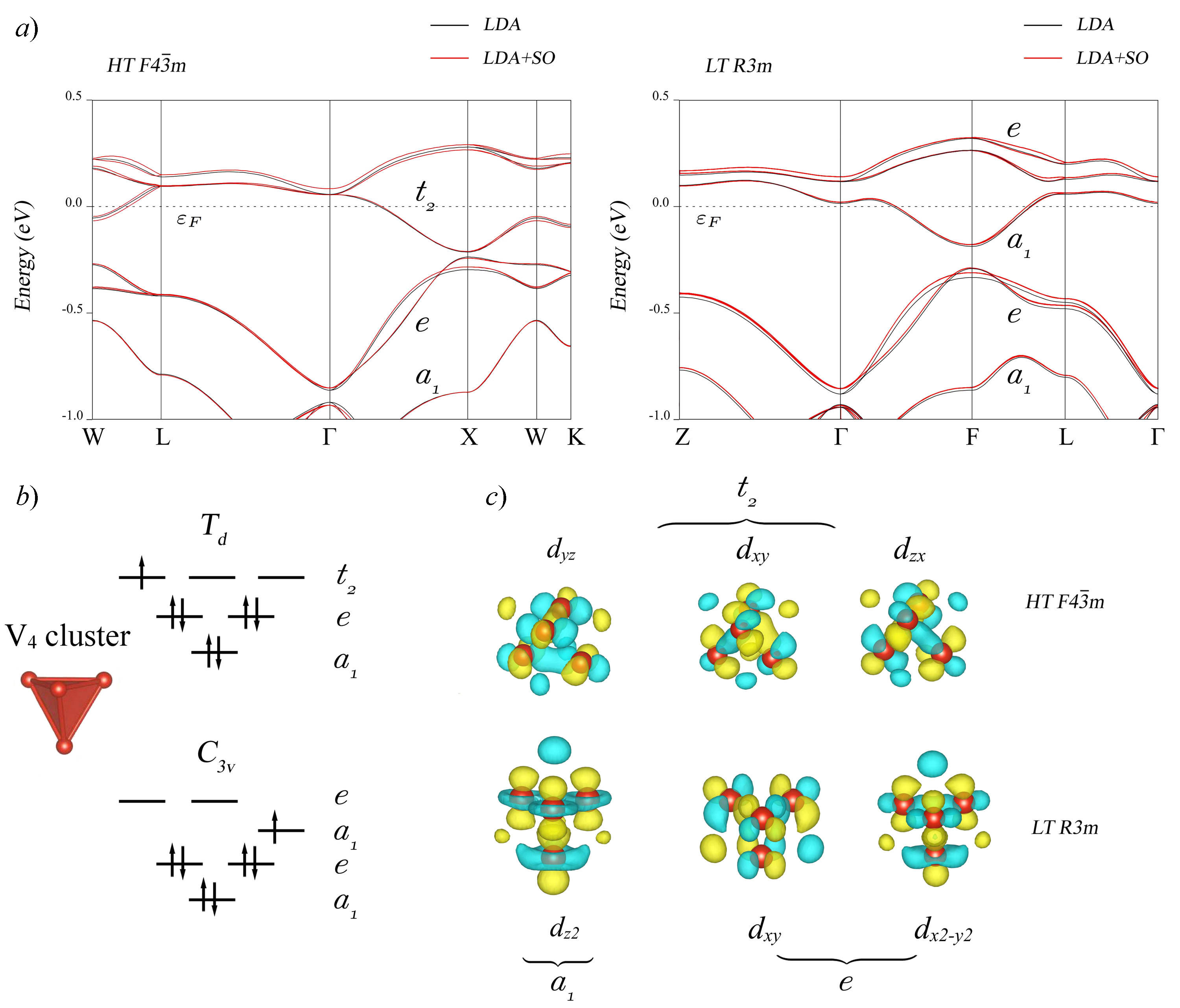}
\end{center}
\caption{$a)$~Bands located near the Fermi level as obtained from LDA and LDA+SO calculations for the high-temperature $F\bar{4}3m$ (left) and low-temperature $R3m$ (right) phases of GaV$_{4}$S$_{8}$; $b)$~Crystal-field splitting of molecular orbitals in the V$_{4}$ cluster; $c)$~Wannier functions as obtained from LDA calculations for the high-temperature $F\bar{4}3m$ (up) and low-temperature $R3m$ (down) phases of GaV$_{4}$S$_{8}$.  Coordinates of the high-symmetry $k$-points are $\mathrm{W}=(1/4,3/4,1/2)$, $\mathrm{L}=(1/2,1/2,1/2)$, $\mathrm{X}=(0,1/2,1/2)$, and $\mathrm{K}=(3/8,3/4,3/8)$  for the $F\bar{4}3m$ space group, and $\mathrm{Z}=(1/2,1/2,1/2)$, $\mathrm{F}=(1/2,1/2,0)$, and $\mathrm{L}=(1/2,0,0)$ for the $R3m$ space group.  }
\label{fig:band}
\end{figure*}

\section{Effective electronic model}
\par The nonmagnetic electronic band structures calculated for the high- and low-temperature phases of GaV$_{4}$S$_{8}$ are presented in Fig.~\ref{fig:band}$a$. The group of bands located near the Fermi level is dominated by metal-metal bonding states of the V$_{4}$ cluster generated by the V $3d$ orbitals. In the high-temperature cubic phase (point group $T_{d}$), these states are split into the $a_{1}$, $e$ and $t_{2}$ levels filled with seven electrons in such a way that the highest 3-fold degenerate $t_{2}$ level contains one unpaired electron. At 38 K GaV$_{4}$S$_{8}$ undergoes  a cubic-to-rhombohedral structural phase transition driven by the cooperative Jahn-Teller distortion \cite{struc} which lifts the degeneracy of the $t_{2}$ level and splits it into the low-lying single $a_{1}$ and double degenerate $e$ states (point group $C_{3v}$), as shown in Fig.~\ref{fig:band}$b$. Since each V$_{4}$ cluster carries a local moment $S=1/2$ formed by three bands located near the Fermi level, these states can be chosen as a basis for the effective low-energy model to describe magnetic properties of GaV$_{4}$S$_{8}$. The corresponding Wannier functions obtained by projecting these states onto atomic-like orbitals are shown in Fig.~\ref{fig:band}$c$. As seen, the resulting Wannier functions are well defined and localized, but have a complex structure representing metal-metal bonding in the V$_{4}$ cluster.
\par For the high-temperature phase, the effective Hubbard-type model can be constructed in the following way:
\begin{equation}
\hat{\mathcal{H}}_{\mathrm{HT}}=\hat{\mathcal{H}}_{\mathrm{kin}}+\hat{\mathcal{H}}_{\mathrm{SO}}+\hat{\mathcal{H}}_{U},
\end{equation}
\noindent the hopping parameters and spin-orbit coupling are given by:
\begin{equation}
\hat{\mathcal{H}}_{\mathrm{kin}}=\sum_{\substack{\langle ij\rangle \\ mm'\sigma}} t_{ij}^{mm'}\hat{c}_{im}^{\dagger\sigma}\hat{c}_{jm'}^{\sigma}
\end{equation}
\noindent and
\begin{equation}
\hat{\mathcal{H}}_{\mathrm{SO}} = \frac{\zeta_{\mathrm{SO}}}{2}\sum_{i,mm'\sigma\sigma'}\hat{c}^{\dagger\sigma}_{im}\left( \begin{array}{ccc}
\phantom{-}0 & -i\sigma_{z} & \phantom{-}i\sigma_{x}\\
\phantom{-}i\sigma_{z} & \phantom{-}0 & -i\sigma_{y} \\
-i\sigma_{x} & \phantom{-}i\sigma_{y} & \phantom{-}0 \\
\end{array} \right)_{mm'}^{\sigma\sigma'}\!\!\!\!\!\!\hat{c}^{\sigma'}_{im'},
\end{equation}
\noindent respectively. Here, $\hat{c}^{\dagger\sigma}_{im} (\hat{c}^{\sigma}_{im})$ is the creation (annihilation) operator of an electron with spin $\sigma$ at site $i$, $m$ is the orbital index in the following order $\{d_{xy},d_{yz},d_{xz}\}$; $t_{ij}^{mm'}$ is the hopping parameter, and $\zeta_{\mathrm{SO}}$ is the spin-orbit coupling constant. For the cubic symmetry, the on-site Coulomb interactions can be parametrized in the Hubbard-Kanamori form \cite{kanamori}:
\begin{equation}
\hat{\mathcal{H}}_{U}=\sum_{i,m}U\hat{n}_{im}^{\uparrow}\hat{n}_{im}^{\downarrow}+\sum_{i,\langle mm' \rangle\sigma}(U-2J)\,\hat{n}_{im}^{\sigma}\hat{n}_{im'}^{\bar{\sigma}}+(U-3J)\,\hat{n}_{im}^{\sigma}\hat{n}_{im'}^{\sigma},
\end{equation}
\noindent where $\hat{n}_{im}^{\sigma}=\hat{c}^{\dagger\sigma}_{im}\hat{c}^{\sigma}_{im}$ is the density operator, $U$ and $U'$ are the intraorbital and interorbital Coulomb interactions, respectively, and $J$ is the interorbital exchange interaction:
\begin{equation*}
U=\frac{1}{N}\sum\limits_{m=1}^{N=3}U^{mmmm},
\end{equation*}
\begin{equation*}
U'=\frac{1}{N(N-1)}\sum\limits_{m\ne m'}^{N=3}U^{mmm'm'},
\end{equation*}
\begin{equation*}
J=\frac{1}{N(N-1)}\sum\limits_{m\ne m'}^{N=3}U^{mm'mm'},
\end{equation*}
\noindent and $U'=U-2J$. Here, we neglect the spin-flip and pair-hopping interactions on account of their smallness. The calculated hopping parameters and Coulomb interactions for the high-temperature phase are given in Tables~II and IV, respectively. The calculated value for the spin-orbit coupling parameter is $\zeta_{\mathrm{SO}}=22.6$ meV.

\begin{table*}[t!]
\caption{Nearest neighbour hopping parameters (in meV) calculated for the $t_{2}$ model of the high-temperature $F\bar{4}3m$ GaV$_{4}$S$_{8}$. The corresponding lattice translations are $\boldsymbol{a}_{1}=a(1,1,0)$, $\boldsymbol{a}_{2}=a(1,0,-1)$, and $\boldsymbol{a}_{3}=a(0,1,-1)$, where $a=4.8305$ \AA.}
\begin{center}
\begin{tabular}{ccc}
\hline
\hline
&& \\
$
\hat{t}_{0\bar{1}1}=\left(
\begin{array}{rrr}
    18.06 &  7.42 & -13.96 \\
     7.42 & 18.06 & 13.96 \\
     13.96 & -13.96 & -36.37
\end{array} \right)
$ &
$
\hat{t}_{01\bar{1}}=\left(
\begin{array}{rrr}
    18.06 &  7.42 & 13.96 \\
     7.42 & 18.06 & -13.96 \\
    -13.96 & 13.96 & -36.37
\end{array} \right)
$ &
$
\hat{t}_{\bar{1}10}=\left(
\begin{array}{rrr}
    18.06 & -13.96 & -7.42 \\
     13.96 & -36.37 & 13.96 \\
   -7.42 & -13.96 & 18.06
\end{array} \right)
$  \\
&&\\
$
\hat{t}_{1\bar{1}0}=\left(
\begin{array}{rrr}
    18.06 &  13.96 & -7.42 \\
    -13.96 & -36.37 & -13.96 \\
    -7.42 & 13.96 & 18.06
\end{array} \right)
$ &
$
\hat{t}_{\bar{1}01}=\left(
\begin{array}{rrr}
    -36.37 &  13.96 & 13.96 \\
    -13.96 & 18.06 & -7.42 \\
    -13.96 & -7.42 & 18.06
\end{array} \right)
$ &
$
\hat{t}_{10\bar{1}}=\left(
\begin{array}{rrr}
     -36.37 & -13.96 & -13.96 \\
     13.96 & 18.06 & -7.42 \\
     13.96 & -7.42 & 18.06
\end{array} \right)
$  \\
&&\\
$
\hat{t}_{0\bar{1}0}=\left(
\begin{array}{rrr}
     -36.37 & -13.96 & 13.96 \\
     13.96 & 18.06 &  7.42 \\
    -13.96 &  7.42 & 18.06
\end{array} \right)
$ &
$
\hat{t}_{010}=\left(
\begin{array}{rrr}
     -36.37 &  13.96 & -13.96 \\
    -13.96 & 18.06 &  7.42 \\
     13.96 &  7.42 & 18.06
\end{array} \right)
$ &
$
\hat{t}_{00\bar{1}}=\left(
\begin{array}{rrr}
     18.06 & 13.96 & 7.42 \\
    -13.96 & -36.37 & 13.96 \\
     7.42 & -13.96 & 18.06
\end{array} \right)
$  \\
&&\\
$
\hat{t}_{001}=\left(
\begin{array}{rrr}
     18.06 & -13.96 & 7.42 \\
     13.96 & -36.37 & -13.96 \\
     7.42 &  13.96 & 18.06
\end{array} \right)
$ &
$
\hat{t}_{\bar{1}00}=\left(
\begin{array}{rrr}
     18.06 & -7.42 & -13.96 \\
    -7.42 & 18.06 & -13.96 \\
     13.96 & 13.96 & -36.37
\end{array} \right)
$ &
$
\hat{t}_{100}=\left(
\begin{array}{rrr}
      18.06 & -7.42 & 13.96 \\
    -7.42 & 18.06 & 13.96 \\
     -13.96 & -13.96 & -36.37
\end{array} \right)
$  \\
&&\\
\hline
\hline
\end{tabular}
\end{center}
\end{table*}

\begin{table*}[t!]
\caption{Coulomb interaction parameters (in eV) calculated for the $t_{2}$ model of the high-temperature $F\bar{4}3m$ GaV$_{4}$S$_{8}$.}
\begin{center}
\begin{tabular}{ccc}
\hline
\hline
Bare Coulomb & Fully screened Coulomb & Partially screened Coulomb\\
\hline
&& \\
$
v^{mmm'm'}=\left(
\begin{array}{ccc}
6.905 & 6.680 & 6.680 \\
6.680 & 6.905 & 6.680 \\
6.680 & 6.680 & 6.905
\end{array} \right)
$ &
$
W^{mmm'm'}=\left(
\begin{array}{ccc}
0.176 & 0.046 & 0.046 \\
0.046 & 0.176 & 0.046 \\
0.046 & 0.046 & 0.176
\end{array} \right)
$ &
$
U^{mmm'm'}=\left(
\begin{array}{ccc}
0.684 & 0.509 & 0.509 \\
0.509 & 0.684 & 0.509 \\
0.509 & 0.509 & 0.684
\end{array} \right)
$  \\
&&\\
$
v^{mm'm'm}=\left(
\begin{array}{ccc}
6.905 & 0.235 & 0.235 \\
0.235 & 6.905 & 0.235 \\
0.235 & 0.235 & 6.905
\end{array} \right)
$ &
$
W^{mm'm'm}=\left(
\begin{array}{ccc}
0.176  & 0.065 & 0.065 \\
0.065 & 0.176  & 0.065 \\
0.065 & 0.065 & 0.176
\end{array} \right)
$ &
$
U^{mm'm'm}=\left(
\begin{array}{ccc}
0.684 & 0.087 & 0.087 \\
0.087 & 0.684 & 0.087 \\
0.087 & 0.087 & 0.684
\end{array} \right)
$  \\
&&\\
$
v^{mm'mm'}=\left(
\begin{array}{ccc}
6.905 & 0.235 & 0.235 \\
0.235 & 6.905 & 0.235 \\
0.235 & 0.235 & 6.905
\end{array} \right)
$ &
$
W^{mm'mm'}=\left(
\begin{array}{ccc}
0.176  & 0.065 & 0.065 \\
0.065 & 0.176  & 0.065 \\
0.065 & 0.065 & 0.176
\end{array} \right)
$ &
$
U^{mm'mm'}=\left(
\begin{array}{ccc}
0.684 & 0.087 & 0.087 \\
0.087 & 0.684 & 0.087\\
0.087 & 0.087 & 0.684
\end{array} \right)
$  \\
&&\\
\multicolumn{3}{c}{Hubbard-Kanamori parameters: $U=0.684$, $U'=0.509$, $J=0.087$ }\\
&&\\
\hline
\hline
\end{tabular}
\end{center}
\end{table*}

\begin{table*}[t]
\caption{Nearest neighbour hopping parameters (in meV) calculated for the $a_{1}\oplus e$ model of the low-temperature $R3m$ GaV$_{4}$S$_{8}$. The corresponding lattice translations are $\boldsymbol{a}_{1}=a(\sqrt{3}/2,-1/2,1.4251)$, $\boldsymbol{a}_{2}=a(0,1,1.4251)$, and $\boldsymbol{a}_{3}=a(-\sqrt{3}/2,-1/2,1.4251)$, where $a=3.9255$ \AA.}
\begin{center}
\begin{tabular}{ccc}
\hline
\hline
&& \\
\multicolumn{3}{c}{Intraplane} \\
&& \\
$
\hat{t}_{\bar{1}01}=\left(
\begin{array}{rrr}
     4.05 & 16.17 &  -25.49 \\
    -16.17 &  10.13 & -18.69 \\
    -25.49 &   18.69 & -10.91 \\
\end{array} \right)
$ &
$
\hat{t}_{10\bar{1}}=\left(
\begin{array}{rrr}
     4.05 & -16.17 &  -25.49 \\
    16.17 &  10.13 & 18.69 \\
    -25.49 &  -18.69 & -10.91 
\end{array} \right)
$ &
$
\hat{t}_{\bar{1}10}=\left(
\begin{array}{rrr}
     4.76 &  29.68 & -1.17 \\
    14.05 & -6.0 & 9.40  \\
     26.52 & -28.44 & 4.51 
\end{array} \right)
$  \\
&&\\
$
\hat{t}_{01\bar{1}}=\left(
\begin{array}{rrr}
     4.76 & -29.68 & -1.17 \\
   -14.05 & -6.0 & -9.40  \\
     26.52 & 28.44 & 4.51 \\
\end{array} \right)
$ &
$
\hat{t}_{1\bar{1}0}=\left(
\begin{array}{rrr}
  4.76 & 14.05 & 26.52 \\
  29.68 & -6.0 & -28.44 \\
  -1.17 & 9.40 &  4.51
\end{array} \right)
$ &
$
\hat{t}_{0\bar{1}1}=\left(
\begin{array}{rrr}
  4.76 & -14.05 & 26.52 \\
 -29.68 & -6.0 & 28.44 \\
  -1.17 & -9.40 &  4.51
\end{array} \right)
$  \\
&& \\
\multicolumn{3}{c}{Interplane} \\
&& \\
$
\hat{t}_{00\bar{1}}=\left(
\begin{array}{rrr}
    -2.70 & 38.02 & 22.18 \\
     0.69 & -8.90 & -18.99 \\
     0.65 &-19.33 & 12.81
\end{array} \right)
$ &
$
\hat{t}_{\bar{1}00}=\left(
\begin{array}{rrr}
    -2.70 & -38.02 & 22.18 \\
    -0.69 & -8.90 & 18.99 \\
     0.65 & 19.33 & 12.81
\end{array} \right)
$ &
$
\hat{t}_{001}=\left(
\begin{array}{rrr}
    -2.70 &  0.69 & 0.65 \\
    38.02 & -8.90 & -19.33 \\
    22.18 & -18.99 & 12.81
\end{array} \right)
$  \\
&&\\
$
\hat{t}_{100}=\left(
\begin{array}{rrr}
    -2.70 &  -0.69 & 0.65 \\
    -38.02 & -8.90 & 19.33 \\
    22.18 & 18.99 & 12.81
\end{array} \right)
$ &
$
\hat{t}_{010}=\left(
\begin{array}{rrr}
    -3.32 & 0.0 & -1.14 \\
    0.0 &  23.97 &  0.0 \\
    -44.28 & 0.0 & -19.45
\end{array} \right)
$ &
$
\hat{t}_{0\bar{1}0}=\left(
\begin{array}{rrr}
    -3.32 & 0.0 & -44.28 \\
    0.0 & 23.97 & 0.0\\
   -1.14 & 0.0 & -19.45
\end{array} \right)
$  \\
&&\\
\hline
\hline
\end{tabular}
\end{center}
\end{table*}

\begin{table*}[t!]
\caption{Coulomb interaction parameters (in eV) calculated for the $a_{1}\oplus e$ model of the low-temperature $R3m$ GaV$_{4}$S$_{8}$.}
\begin{center}
\begin{tabular}{ccc}
\hline
\hline
Bare Coulomb & Fully screened Coulomb & Partially screened Coulomb\\
\hline
&& \\
$
v^{mmm'm'}=\left(
\begin{array}{ccc}
7.177 & 6.556 & 6.553 \\
6.556 & 6.992 & 6.630 \\
6.553 & 6.630 & 6.995
\end{array} \right)
$ &
$
W^{mmm'm'}=\left(
\begin{array}{ccc}
0.159 &  0.048 &     0.048 \\
0.048 &  0.216 &     0.073 \\
0.048 &  0.073 &     0.216 \\
\end{array} \right)
$ &
$
U^{mmm'm'}=\left(
\begin{array}{ccc}
0.707 & 0.509 & 0.509 \\
0.509 & 0.676 & 0.505 \\
0.509 & 0.505 & 0.676
\end{array} \right)
$  \\
&&\\
$
v^{mm'm'm}=\left(
\begin{array}{ccc}
 7.177 & 0.139 & 0.136 \\
0.139 &  6.992 & 0.179 \\
0.136 & 0.179 & 6.995
\end{array} \right)
$ &
$
W^{mm'm'm}=\left(
\begin{array}{ccc}
0.152 &   0.062 &     0.062 \\
0.062 &   0.216 &     0.071 \\
0.062 &   0.071 &     0.216
\end{array} \right)
$ &
$
U^{mm'm'm}=\left(
\begin{array}{ccc}
0.707 & 0.086 &  0.086 \\
0.086 & 0.676 &  0.086 \\
0.086 & 0.086 &  0.676
\end{array} \right)
$  \\
&&\\
$
v^{mm'mm'}=\left(
\begin{array}{ccc}
 7.177 & 0.139 & 0.136 \\
0.139 &  6.992 & 0.179 \\
0.136 & 0.179 & 6.995
\end{array} \right)
$ &
$
W^{mm'mm'}=\left(
\begin{array}{ccc}
0.152 &  0.062 &  0.062 \\
0.062 &  0.216 &  0.071 \\
0.062 &  0.071 &  0.216 \\
\end{array} \right)
$ &
$
U^{mm'mm'}=\left(
\begin{array}{ccc}
0.707 & 0.086 &  0.086 \\
0.086 & 0.676 &  0.086 \\
0.086 & 0.086 &  0.676
\end{array} \right)
$  \\
&&\\
\hline
\hline
\end{tabular}
\end{center}
\end{table*}

\par The corresponding Hubbard-type model for the $a_{1}\oplus e$ states of the low-temperature phase has the following form:
\begin{equation}
\hat{\mathcal{H}}_{\mathrm{LT}}=\hat{\mathcal{H}}_{\mathrm{kin}}+\hat{\mathcal{H}}_{\mathrm{CF}}+\hat{\mathcal{H}}_{\mathrm{SO}}+\hat{\mathcal{H}}_{\mathrm{U}},
\label{eq:lowhub}
\end{equation}
\noindent  the hopping parameters and crystal-field splitting are given by:
\begin{equation}
\hat{\mathcal{H}}_{\mathrm{kin}}=\sum_{\substack{\langle ij\rangle \\ mm'\sigma}} t_{ij}^{mm'}\hat{c}_{im}^{\dagger\sigma}\hat{c}_{jm'}^{\sigma}
\end{equation}
\noindent and
\begin{equation}
\hat{\mathcal{H}}_{\mathrm{CF}}=\sum_{i,m\in e,\sigma} \Delta\,\hat{c}_{im}^{\dagger\sigma}\hat{c}_{im}^{\sigma},
\end{equation}
\noindent where $m$ runs over the orbitals in the following order $\{d_{z^{2}},d_{xy},d_{x^{2}-y^{2}} \}$; $\Delta$ is the crystal-field splitting between the $a_{1}$ and $e$ orbitals. The spin-orbit coupling term is expressed as:
\begin{equation}
\hat{\mathcal{H}}_{\mathrm{SO}} = \frac{\zeta_{\mathrm{SO}}}{2}\sum_{i,mm'\sigma\sigma'}\hat{c}^{\dagger\sigma}_{im}\left( \begin{array}{ccc}
\phantom{-}0 & -i\sigma_{y} & \phantom{-}i\sigma_{x}\\
\phantom{-}i\sigma_{y} & \phantom{-}0 & \phantom{-}i\sigma_{z} \\
-i\sigma_{x} & -i\sigma_{z} & \phantom{-}0 \\
\end{array} \right)_{mm'}^{\sigma\sigma'}\!\!\!\!\!\!\hat{c}^{\sigma'}_{m'} - \frac{\zeta_{\mathrm{SO}}^{R}}{2}\sum_{i,mm'\sigma\sigma'}\hat{c}^{\dagger\sigma}_{im}\left(\begin{array}{ccc}
\phantom{-}0 & -i\sigma_{y} & \phantom{-}i\sigma_{x}\\
\phantom{-}i\sigma_{y} & \phantom{-}0 & \phantom{-}0 \\
-i\sigma_{x} &\phantom{-}0 & \phantom{-}0 \\
\end{array} \right)_{mm'}^{\sigma\sigma'}\!\!\!\!\!\!\hat{c}^{\sigma'}_{im'},
\end{equation}
\noindent where the second term comes from the Jahn-Teller distortion of the V$_{4}$ cluster along the $z$ axis and has a Rashba-type form. The first (spherical) term can be presented in the form $\zeta_{\mathrm{SO}}\hat{\boldsymbol{L}}\cdot\hat{\boldsymbol{S}}$, which can be used to derive matrix elements of the angular momentum operator $\hat{\boldsymbol{L}}$ in the Wannier basis. The matrix elements of the angular momentum operators can be written in a compact form in terms of the Levi-Civita symbol, $( \hat{L}^{x}_{i} )^{ab} = -i\varepsilon_{2ab}$, $( \hat{L}^{y}_{i} )^{ab} = -i\varepsilon_{3ab}$, and $( \hat{L}^{z}_{i} )^{ab} = i\varepsilon_{1ab}$. Finally, the Hubbard term is written as:
\begin{equation}
\hat{\mathcal{H}}_{\mathrm{U}}=\sum_{i,mm'\sigma}U_{}^{mm'}\hat{n}_{im}^{\sigma}\hat{n}_{im}^{\bar{\sigma}}+\sum_{i,\langle mm' \rangle\sigma}(U^{mm'}_{}-J^{mm'}_{})\hat{n}_{im}^{\sigma}\hat{n}_{im'}^{\sigma}
\label{eq:hublow}
\end{equation}
\noindent where $U^{mm'}=U^{mmm'm'}$ and $J^{mm'}=U^{mm'm'm}$ are the on-site Coulomb and exchange interactions. The calculated hopping parameters and Coulomb interactions for the low-temperature phase are presented in Tables~III and V, respectively. The calculated value of the crystal-field splitting is $\Delta=98.1$ meV. The spin-orbit coupling parameters are $\zeta_{SO}=23.1$ meV and $\zeta_{SO}^{R}=1.2$ meV.  Finally, on-site matrix elements of the position operator in the constructed basis are $(\hat{r}^{z})^{11}=7.5329$ \AA, $(\hat{r}^{z})^{22}=(\hat{r}^{z})^{33}=7.3772$ \AA, and $(\hat{r}^{x})^{ij}=(\hat{r}^{y})^{ij}\approx0$.

\par According to our results, the on-site Coulomb interactions are not particularly strong but still remain the largest parameters in the model. This can be understood as follows. First, the resulting Wannier orbitals are rather extended (the calculated spread is $\sim8.5$ \AA$^{2}$). Therefore, the bare Coulomb and exchange interactions in the Wannier basis, $U = 6.91$ eV and $J = 0.24$ eV, are already considerably reduced compared to their atomic values for real $d$ orbitals, $U^{A}$ and $J^{A}$. Since the interatomic exchange integrals are small, the main contribution to $J$ is due to the bare intraatomic interaction, $J^{A} \sim 1$ eV. Assuming that the Wannier orbitals are equally distributed over the four atomic sites in the $V_{4}$ cluster, $J$ can be evaluated as $J \sim \frac{1}{4}J^{A}$, which is consistent with the calculated value. Second, the bare interactions are efficiently screened due to the proximity of the target $t_{2}$ bands to the low-lying $a_{1}$ and $e$ states of the V$_{4}$ cluster.

\section{Hartree-Fock approximation}
\par The effective low-energy models allow us to go beyond conventional first-principles calculations that often fail to describe system with electronic correlations. In particular, they come in handy when dealing with ``exotic" systems, like GaV$_{4}$S$_{8}$, where the properties of interest are featured by complex molecular orbitals, so that the use of ordinary methods implemented for atomic-like orbitals (for example, LDA+U) may be questionable.
\par There is a large variety of different techniques used to deal with the Hubbard model:
\begin{equation}
\begin{aligned}
\hat{\mathcal{H}}=\sum_{ij \sigma\sigma'}\sum_{ \alpha\beta}t_{ij}^{\alpha\beta}\hat{c}^{\dagger\sigma}_{i\alpha}\hat{c}^{\sigma'}_{j\beta}&+\frac{1}{2}\sum_{i\sigma\sigma'}\sum_{\alpha\beta\gamma\delta}U_{i}^{\alpha\beta\gamma\delta}\hat{c}^{\dagger\sigma}_{i\alpha}\hat{c}^{\dagger\sigma'}_{i\gamma}\hat{c}^{\sigma}_{i\beta}\hat{c}^{\sigma'}_{i\delta}.
\end{aligned}
\end{equation}
\noindent In the simplest way, this model can be solved by expressing the correlation term as:
\begin{equation}
\begin{aligned}
\hat{c}^{\dagger\sigma}_{i\alpha}\hat{c}^{\dagger\sigma'}_{i\gamma}\hat{c}^{\sigma}_{i\beta}\hat{c}^{\sigma'}_{i\delta}&\approx \hat{c}^{\dagger\sigma}_{i\alpha}\hat{c}^{\sigma}_{i\beta}\langle\hat{c}^{\dagger\sigma'}_{i\gamma}\hat{a}^{\sigma'}_{i\delta}\rangle + \hat{c}^{\dagger\sigma'}_{i\gamma}\hat{c}^{\sigma'}_{i\delta}\langle\hat{c}^{\dagger\sigma}_{i\alpha}\hat{c}^{\sigma}_{i\beta}\rangle \\
&- \hat{c}^{\dagger\sigma}_{i\alpha}\hat{c}^{\sigma'}_{i\delta}\langle\hat{c}^{\dagger\sigma'}_{i\gamma}\hat{c}^{\sigma}_{i\beta}\rangle - \hat{c}^{\dagger\sigma'}_{i\gamma}\hat{c}^{\sigma}_{i\beta}\langle\hat{c}^{\dagger\sigma}_{i\alpha}\hat{c}^{\sigma'}_{i\delta}\rangle,
\end{aligned}
\end{equation}
\noindent where all higher-order quantum fluctuations are neglected. Here, $\langle...\rangle$ stands for an average over the ground state. By using symmetry properties of the Coulomb matrix, we obtain the following one-electron Hamiltonian \cite{hfsol}:
\begin{figure*}
\begin{center}
\includegraphics[width=0.80\textwidth]{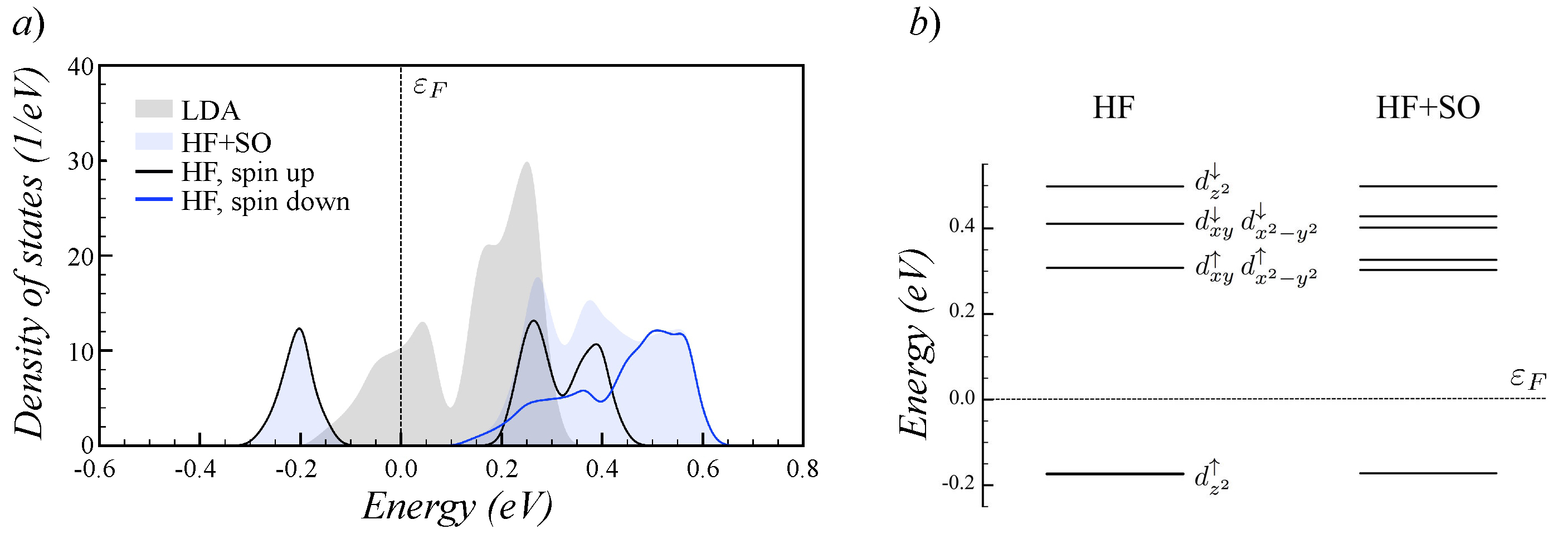}
\end{center}
\caption{$a$) Density of states  and $b$) the on-site level splitting of the ferromagnetic state in the low-temperature phase as obtained in the Hartree-Fock approximation with and without spin-orbit coupling (HF+SO and HF, respectively).}
\label{fig:dos}
\end{figure*}
\begin{equation}
\hat{\mathcal{H}}^{\mathrm{HF}}=\sum_{ij\sigma\sigma'}\sum_{\alpha\beta}\left(t_{ij}^{\alpha\beta}\delta^{}_{\sigma\sigma'}+\mathcal{V}_{i}^{\alpha\beta\sigma\sigma'}\delta^{}_{ij}\right)\hat{c}^{\dagger\sigma}_{i\alpha}\hat{c}^{\sigma'}_{j\beta},
\end{equation}
\noindent where
\begin{equation}
\mathcal{V}_{i}^{\alpha\beta\sigma\sigma'}=\sum_{\gamma\delta}U_{i}^{\alpha\beta\gamma\delta}(n_{i\gamma\delta}^{\uparrow\uparrow}+n_{i\gamma\delta}^{\downarrow\downarrow})\delta_{\sigma\sigma'}-U_{i}^{\alpha\delta\gamma\beta} n_{i\gamma\delta}^{\sigma'\sigma}
\end{equation}
\noindent or
\begin{equation}
\hat{\mathcal{V}}_{i} = \hat{\mathbb{U}}_{i}\hat{n}_{i} \qquad 
\left(\begin{array}{c}
\mathcal{V}_{i}^{\alpha\beta\uparrow\uparrow} \\ \mathcal{V}_{i}^{\alpha\beta\downarrow\uparrow} \\ \mathcal{V}_{i}^{\alpha\beta\uparrow\downarrow} \\ \mathcal{V}_{i}^{\alpha\beta\downarrow\downarrow} 
\end{array}\right) = \sum_{\gamma\delta}
\left(\begin{array}{cccc}
U_{i}^{\alpha\beta\gamma\delta}-U_{i}^{\alpha\delta\gamma\beta} & 0 & 0 & U_{i}^{\alpha\beta\gamma\delta} \\
0 & 0 & -U_{i}^{\alpha\delta\gamma\beta} & 0 \\
0 & -U_{i}^{\alpha\delta\gamma\beta} & 0 & 0 \\
U_{i}^{\alpha\beta\gamma\delta} & 0 & 0 & U_{i}^{\alpha\beta\gamma\delta}-U_{i}^{\alpha\delta\gamma\beta} \\
\end{array}\right)
\left(\begin{array}{c}
n_{i\gamma\delta}^{\uparrow\uparrow} \\ n_{i\gamma\delta}^{\downarrow\uparrow} \\ n_{i\gamma\delta}^{\uparrow\downarrow} \\ n_{i\gamma\delta}^{\downarrow\downarrow} \\
\end{array}\right),
\end{equation}
\noindent and
\begin{equation}
n_{i\gamma\delta}^{\sigma'\sigma}=\langle\hat{c}^{\dagger\sigma'}_{i\gamma}\hat{c}^{\sigma}_{i\delta}\rangle.
\end{equation}
\noindent In the reciprocal space, it is equivalent to solving the Hartree-Fock equations:
\begin{equation}
\big(\hat{t}_{\boldsymbol{k}}-\hat{\mathcal{V}} \big)|\varphi_{n\boldsymbol{k}}\rangle=\varepsilon_{n\boldsymbol{k}}|\varphi_{n\boldsymbol{k}}\rangle,
\end{equation}
\noindent where $\hat{t}_{\boldsymbol{k}}$ is the Fourier image of $\hat{t}_{ij}$, $|\varphi_{n\boldsymbol{k}}\rangle=[\varphi_{n\boldsymbol{k}}^{i\alpha\sigma}]$ is an eigenvector containing site $i$, orbital $\alpha$, and spin $\sigma$ indices. An iterative solution of the Hartree-Fock equations is obtained self-consistently with respect to the total energy:
\begin{equation}
E^{\mathrm{HF}}=\sum_{\boldsymbol{k}}\sum_{n}^{occ}\varepsilon_{n\boldsymbol{k}}-\frac{1}{2}\sum\limits_{i}\mathrm{Tr}\,\big\{\hat{n}_{i}^{T}\hat{\mathcal{V}}_{i}^{}\big\}.
\end{equation}
\par The results of Hartree-Fock calculations are shown in Fig.~\ref{fig:dos}.

\section{Spin model}
\par For the low-temperature phase of GaV$_{4}$S$_{8}$, the electronic model, Eq.~(\ref{eq:lowhub}), can be mapped onto the classical spin model:
\begin{equation}
\mathcal{H}^{\mathrm{S}} = -\sum\limits_{\langle ij\rangle} J_{ij}\boldsymbol{e}_{i}\cdot\boldsymbol{e}_{j}+\sum\limits_{\langle ij\rangle} \boldsymbol{D}_{ij}\cdot\boldsymbol{e}_{i}\times\boldsymbol{e}_{j},
\label{eq:spinmodel}
\end{equation}
\noindent where the first and second terms stand for the isotropic Heisenberg and anisotropic Dzyaloshinskii-Moriya exchange interactions, respectively, and $\boldsymbol{e}_{i}$ is the unit magnetic moment at site $i$. Below, we consider two approaches for calculating exchange parameters.

\subsection{The Green's functions approach}
\par The exchange parameters can be derived from a collinear configuration obtained in Hartree-Fock calculations without spin-orbit coupling by considering a perturbation $\delta\hat{v}^{p}_{i}$ caused by applying the spin-orbit coupling potential at one site while rotating the magnetic moment at another site. The corresponding rotation is given by another perturbation, $\delta\hat{v}^{r}_{j}=\hat{b}_{j}\delta\boldsymbol{e}_{j}\cdot\hat{\boldsymbol{\sigma}}$, where $\hat{b}_{j}=\frac{1}{2}(\hat{\mathcal{V}}^{\uparrow}_{j}-\hat{\mathcal{V}}^{\downarrow}_{j})$ is the exchange splitting field at site $j$. Having employed Lloyd's formula \cite{lloyd} to express the change in the single-particle energy to second order with respect to the applied potentials, the exchange parameters are obtained from the force $\boldsymbol{f}_{i}=-\partial E/\partial\boldsymbol{e}_{i}$ rotating the magnetic moment at site $i$, which for the spin model given above is written as \cite{solresp}:
\begin{equation}
\boldsymbol{f}_{i}=\sum_{j}\boldsymbol{f}_{i}^{j}=\frac{1}{2}\sum_{j}\big(J_{ij}\boldsymbol{e}_{j}+\boldsymbol{D}_{ij}\times\boldsymbol{e}_{j}\big),
\end{equation}
\noindent where $\boldsymbol{f}_{i}^{j}$ is the contribution to the force from site $j$, and the exchange interactions are derived from:
\begin{equation}
J_{ij}=\frac{2}{\pi}\int\limits_{\infty}^{\varepsilon_{F}}dE\,\mathrm{Tr}_{L}\{\hat{G}_{ij}^{\uparrow}(E)\hat{b}_{j}\hat{G}_{ji}^{\downarrow}(E)\hat{b}_{i}  \},
\end{equation}
\begin{equation}
\boldsymbol{f}^{j}_{i}-\boldsymbol{f}^{i}_{j}=\frac{1}{\pi}\int\limits_{\infty}^{\varepsilon_{F}}dE\,\mathrm{Tr}_{L}\Big\{\hat{\boldsymbol{\sigma}}\big(\hat{b}_{i}\hat{\tilde{G}}_{ij}(E)\delta\hat{v}^{p}_{j}\hat{\tilde{G}}_{ji}(E)-\hat{b}_{j}\hat{\tilde{G}}_{ji}(E)\delta\hat{v}^{p}_{i}\hat{\tilde{G}}_{ij}(E)\big) \Big\},
\end{equation}
\noindent where the unperturbed Green's function is defined as:
\begin{equation}
\hat{\tilde{G}}_{ij}=\left(
\begin{array}{cc}
\hat{G}^{\uparrow}_{ij} & 0 \\
0 & \hat{G}^{\downarrow}_{ij}
\end{array}\right)\qquad\qquad
\hat{G}^{\sigma}_{ij}(E)=\sum_{n\boldsymbol{k}}\frac{|\varphi^{\sigma}_{n\boldsymbol{k}}\rangle\langle\varphi^{\sigma}_{n\boldsymbol{k}}|}{E-\varepsilon_{n\boldsymbol{k}}+i\delta}e^{i\boldsymbol{k}\cdot(\boldsymbol{R}_{i}-\boldsymbol{R}_{j})}.
\end{equation}
\noindent By considering a small deviation of the magnetic moment at each site, $\boldsymbol{e}_{i}=\boldsymbol{e}_{0}+\delta\boldsymbol{e}_{i}$, the corresponding Dzyaloshinskii-Moriya interactions can be derived from:
\begin{equation}
\boldsymbol{f}^{j}_{i}-\boldsymbol{f}^{i}_{j}=\boldsymbol{D}\times\boldsymbol{e}_{0}+\frac{1}{2}J_{ij}(\delta\boldsymbol{e}_{j}-\delta\boldsymbol{e}_{i}).
\end{equation}

\begin{figure*}[t!]
\begin{center}
\includegraphics[width=0.90\textwidth]{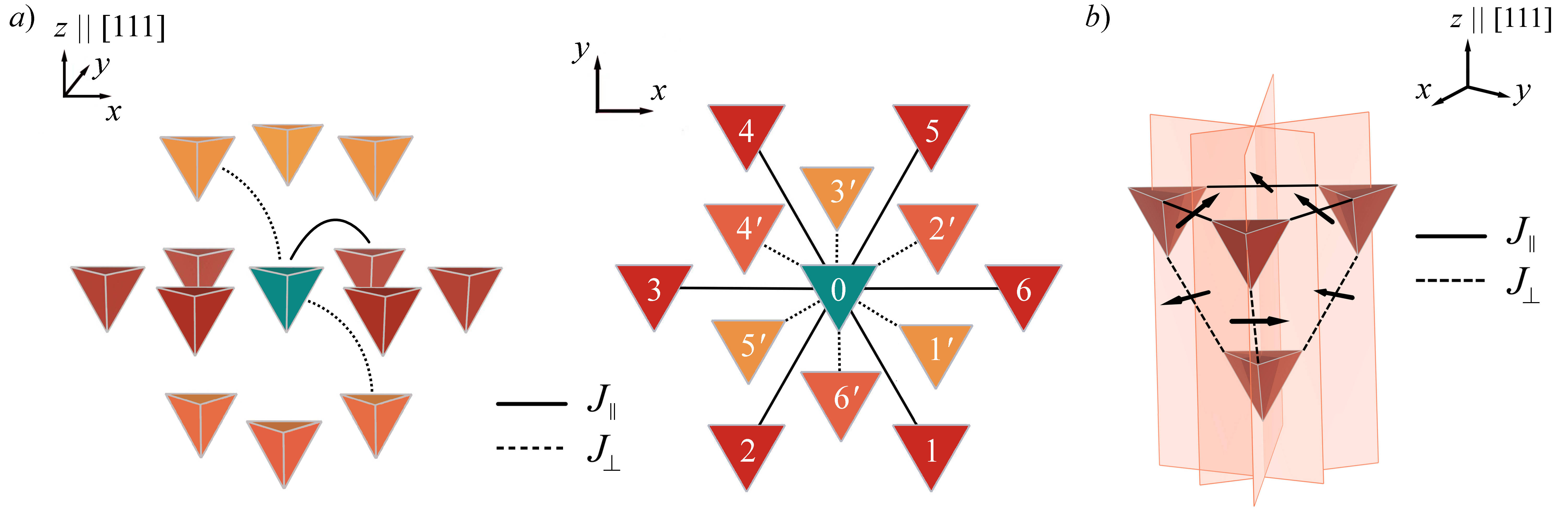}
\end{center}
\caption{$a)$~Spin model for the low-temperature GaV$_{4}$S$_{8}$ shown along the $z\parallel[111]$ direction; $b)$~Sketch view of the mirror planes of the $C_{3v}$ point group. The Dzyaloshinskii-Moriya exchange interactions are schematically shown with black arrows.}
\label{fig:spin}
\end{figure*}

\begin{table*}[t!]
\caption{Exchange parameters (in meV) for the spin model, Eq.~(\ref{eq:spinmodel}), as obtained from the mean-field Hartree-Fock calculations (HF) and superexchange theory (SE). See Fig.~\ref{fig:spin}$a$ for details.}
\begin{center}
\begin{tabular}{c|c}
\hline
\hline
HF & SE  \\
\hline
\multicolumn{2}{c}{Intraplane}\\ 
$J_{\parallel}=0.202$ & $J_{\parallel}=0.180$ \\
$
\begin{aligned}
\boldsymbol{D}_{01}&=(0.138,0.080,0.010) \\
\boldsymbol{D}_{02}&=(0.138,-0.080,-0.010) \\
\boldsymbol{D}_{03}&=(0.0,-0.159,0.010) \\
\boldsymbol{D}_{04}&=(-0.138,-0.080, -0.010) \\
\boldsymbol{D}_{05}&=(-0.138,0.080,0.010) \\
\boldsymbol{D}_{06}&=(0.0,0.159,-0.010)  \\
\end{aligned}
$
&
$
\begin{aligned}
\boldsymbol{D}_{01}&=(0.063,0.036,-0.010)\\
\boldsymbol{D}_{02}&=(0.063,-0.036,0.010)\\
\boldsymbol{D}_{03}&=(0.0,-0.073,-0.010)\\
\boldsymbol{D}_{04}&=(-0.063,-0.036,0.010) \\
\boldsymbol{D}_{05}&=(-0.063,0.036,-0.010)\\
\boldsymbol{D}_{06}&=(0.0,0.073,0.010) \\
\end{aligned}
$\\
\multicolumn{2}{c}{Interplane}\\ 
$J_{\perp}=0.363$ & $J_{\perp}=0.217$ \\
$
\begin{aligned}
\boldsymbol{D}_{01'}&=( -0.021,-0.036,0.0)  \\
\boldsymbol{D}_{02'}&=(0.021,-0.036,0.0) \\
\boldsymbol{D}_{03'}&=(0.041,0.0,0.0) \\
\boldsymbol{D}_{04'}&=(0.021,0.036,0.0) \\
\boldsymbol{D}_{05'}&=(-0.021,0.036,0.0) \\
\boldsymbol{D}_{06'}&=(-0.041,0.0,0.0) \\
\end{aligned}
$
&
$
\begin{aligned}
\boldsymbol{D}_{01'}&=(0.029,0.049,0.0) \\
\boldsymbol{D}_{02'}&=( -0.029,0.049,0.0)\\
\boldsymbol{D}_{03'}&=(-0.057,0.0,0.0)\\
\boldsymbol{D}_{04''}&=(-0.029,-0.049,0.0)\\
\boldsymbol{D}_{05'}&=( 0.029,-0.049,0.0) \\
\boldsymbol{D}_{06'}&=( 0.057,0.0,0.0)\\
\end{aligned}
$\\
\hline
\hline
\end{tabular}
\end{center}
\end{table*}

\par The effect of spin-orbit coupling is included perturbatively within the self-consistent linear response theory~\cite{solresp}, where the change of the density matrix with respect to an external perturbation $\delta\hat{v}^{\mathrm{ext}}_{i}=\hat{\mathcal{H}}_{\mathrm{SO}}$ is given by:
\begin{equation}
\delta\hat{n}_{i}=\hat{\mathcal{R}}_{ij}\delta\hat{v}^{\mathrm{ext}}_{j}
\end{equation}
\noindent with the corresponding change of the Hartree-Fock potential:
\begin{equation}
\delta \hat{v}_{i} =\hat{\mathbb{U}}_{i}\delta \hat{n}_{i},  
\end{equation}
\noindent where the response matrix $\hat{\mathcal{R}}_{ij}=[\mathcal{R}_{ij,\alpha\beta\gamma\delta}^{\sigma\sigma'}]$ is defined as:
\begin{equation}
\mathcal{R}_{ij,\alpha\beta\gamma\delta}^{\sigma\sigma'}=\sum\limits_{n}^{occ}\sum\limits_{n'}^{unocc}\sum_{\boldsymbol{k}}\left(\frac{(\varphi_{n\boldsymbol{k}}^{i\alpha\sigma})^{*}\varphi_{n'\boldsymbol{k}}^{i\beta\sigma'}(\varphi_{n'\boldsymbol{k}}^{j\gamma\sigma'})^{*}\varphi_{n\boldsymbol{k}}^{j\delta\sigma'}}{\varepsilon_{n\boldsymbol{k}}^{\sigma}-\varepsilon_{n'\boldsymbol{k}}^{\sigma'}} + \frac{(\varphi_{n'\boldsymbol{k}}^{i\alpha\sigma})^{*}\varphi_{n\boldsymbol{k}}^{i\beta\sigma'}(\varphi_{n\boldsymbol{k}}^{j\gamma\sigma'})^{*}\varphi_{n'\boldsymbol{k}}^{j\delta\sigma'} }{\varepsilon_{n\boldsymbol{k}}^{\sigma'}-\varepsilon_{n'\boldsymbol{k}}^{\sigma}}    \right).
\end{equation}
\noindent Combining $\delta\hat{v}_{i}$ with $\delta\hat{v}^{\mathrm{ext}}_{i}$, the problem is solved iteratively to give the total perturbation $\delta\hat{v}^{p}_{i}=\delta\hat{v}^{\mathrm{ext}}_{i}+\delta\hat{v}_{i}$~\cite{solresp}:
\begin{equation}
\delta\hat{v}^{p}_{i} = \big(1 - \hat{\mathbb{U}}_{i}\hat{\mathbb{R}}_{ij}\big)^{-1}\delta\hat{v}^{\mathrm{ext}}_{j} \qquad \qquad \delta\hat{n}_{i}=\hat{\mathbb{R}}_{ij}\delta\hat{v}^{p}_{j}
\end{equation}
\noindent with
\begin{equation}
\hat{\mathbb{R}}_{ij}=
\left(\begin{array}{cccc}
\hat{\mathcal{R}}^{\uparrow\uparrow}_{ij} & 0 & 0 & 0 \\
0 & 0 & \hat{\mathcal{R}}^{\downarrow\uparrow}_{ij}  & 0 \\
0 & \hat{\mathcal{R}}^{\uparrow\downarrow}_{ij}  & 0 & 0 \\
0 & 0 & 0 & \hat{\mathcal{R}}^{\downarrow\downarrow}_{ij}  \\
\end{array}\right).
\end{equation}
\par The calculated exchange parameters are given in Table VI. As can be seen from Fig.~\ref{fig:spin}$b$, these parameters obey the $C_{3v}$ point group and are in agreement with Moriya rules. Clearly, because the mirror planes pass along the bonds connecting nearest neighbours in the adjacent $[111]$ planes, the anisotropic exchange parameters should be perpendicular to the corresponding bonds and lie in the $xy$ plane. At the same time, the mirror planes are perpendicular to the in-plane bonds connecting nearest neighbours, and the corresponding Dzyaloshinskii-Moriya interactions lie within the plane with an alternating $z$ component.

\subsection{Superexchange theory}
\par When $t/U\ll1$, we can start from the atomic limit:
\begin{equation}
\hat{\mathcal{H}}_{0} = \hat{\mathcal{H}}_{\mathrm{CF}}+\hat{\mathcal{H}}_{\mathrm{SO}}+ \hat{\mathcal{H}}_{U}\qquad \hat{\mathcal{H}}_{U}=\frac{1}{2}U^{\alpha\beta\gamma\delta}\hat{c}_{\alpha}^{\dagger}\hat{c}_{\gamma}^{\dagger}\hat{c}_{\beta}^{\phantom{\dagger}}\hat{c}_{\delta}^{\phantom{\dagger}},
\end{equation}
\noindent and consider hopping parameters as a small perturbation. Superexchange interaction is regarded as the kinetic energy gain acquired by an occupied electron centering at one site in the process of virtual hoppings into the subspace of unoccupied states at neighboring sites (and vice versa) \cite{Anderson,KugelKhomskii}. In the atomic limit, this energy gain can be obtained to second order in hopping parameters as \cite{solse1,solse2}:
\begin{equation}
\tensor{\mathcal{T}}_{ij} = - \left\langle \mathcal{G}_{ij}\left| \hat{t}_{ij} \left( \sum_M \frac{{\hat{\mathcal{P}}}_{j} |jM
\rangle \langle jM| \hat{\mathcal{P}}_{j}}{E_{jM}} \right) \hat{t}_{ji} + (i \leftrightarrow {j}) \right| \mathcal{G}_{ij} \right\rangle.
\label{eq:seexchnage}
\end{equation}
\noindent  This version of superexchange theory combines the one-electron approximation and many-body effects and in this sense goes beyond the conventional mean-field approximation. Here, $\mathcal{G}_{ij}$ is the ground-state wavefunction for the bond $ij$. Since each atomic site accommodates one electron, one can start from the one-electron approximation and express ${\cal G}_{ij}$ in the form of a Slater determinant constructed from the occupied orbitals $\phi_{i}$ at sites $i$ and $j$:
\begin{equation}
\mathcal{G}_{ij}(1,2) = \frac{1}{\sqrt{2}}\big[\phi_{i}(1)\phi_{j}(2)-\phi_{i}(2)\phi_{j}(1)\big].
\end{equation}
\noindent The orbitals $\phi_{i}$ correspond to the lowest Kramers doublet obtained by diagonalising $\hat{\mathcal{H}}_{\mathrm{CF}}+\hat{\mathcal{H}}_{\mathrm{SO}}$ and are taken as their linear combinations to diagonalise the Pauli matrices (satisfying the maximal projection of the pseudospin moment in a given direction $\boldsymbol{e}_{i}$). 
\begin{figure*}[t!]
\begin{center}
\includegraphics[width=0.6\textwidth]{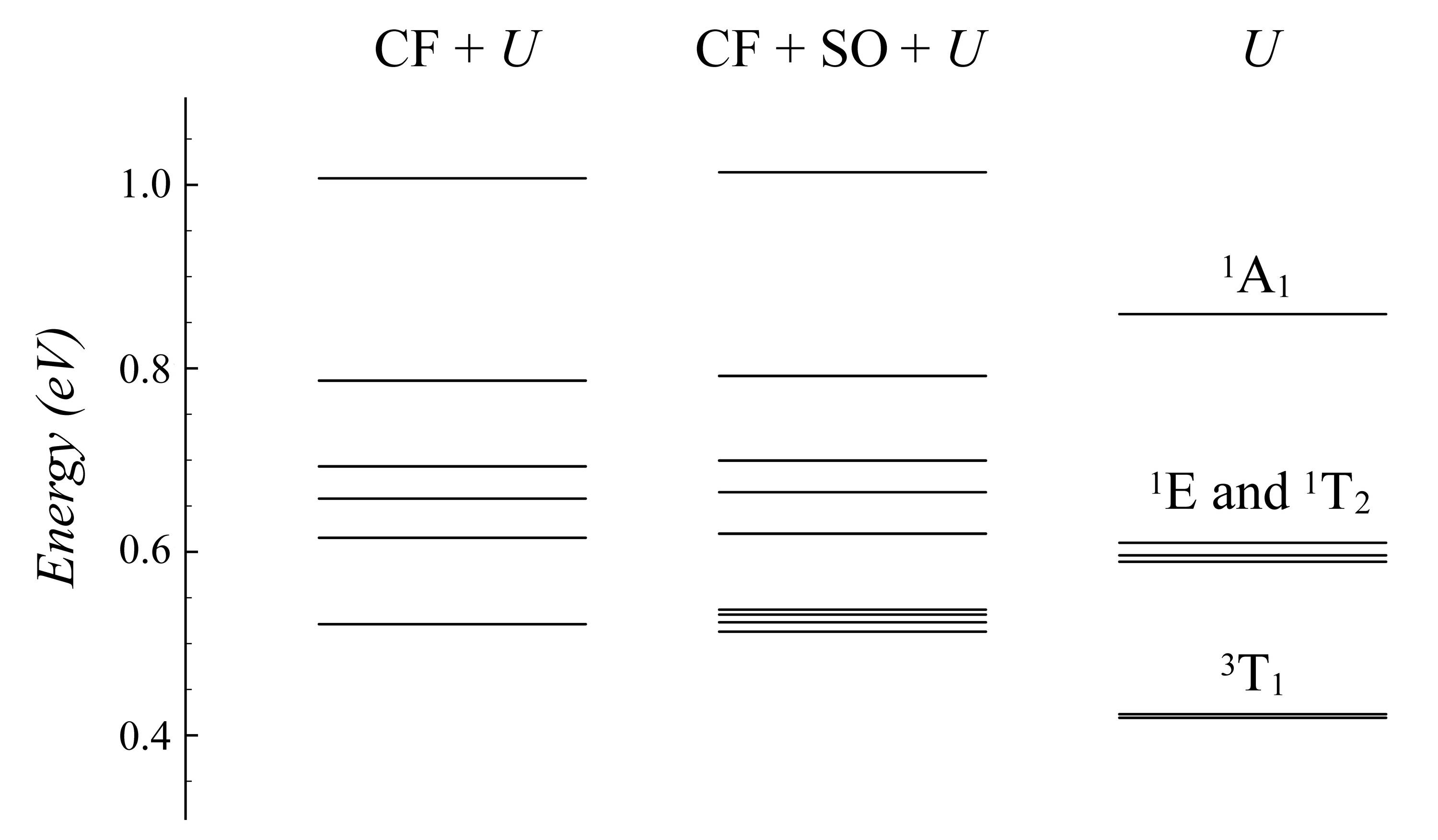}
\end{center}
\caption{Energies of the excited two-electron states for the low-temperature $R3m$ GaV$_{4}$S$_{8}$ obtained in the atomic limit by taking into account crystal-field splitting (CF), spin-orbit coupling (SO), and on-site Coulomb interactions ($U$).}
\label{fig:2e}
\end{figure*}
\par Next, we consider virtual hoppings of the occupied electron $\phi_{i}$ at site $i$ to the unoccupied two-electron states at neighbouring sites $j$. Here, $E_{jM}$ and $|jM\rangle$ stand for the true eigenvalues and eigenvectors of the excited two-electron configurations, which are obtained in the basis of all possible Slater determinants at site $j$ (for the $t_{2}$ model, each $|jM \rangle$ is expressed in the basis of $6\times5/2 = 15$ Slater determinants). The corresponding two-electron Hamiltonian $\hat{\mathcal{H}}_{\mathrm{2e}}$ in the basis of Slater determinants is constructed from $\hat{\mathcal{H}}_{0}$ by using Slater-Condon rules:
\begin{equation}
\mathcal{H}^{nm}_{\mathrm{2e}} = U^{i1,i2,i3,i4}-U^{i1,i4,i3,i2} + \mathcal{H}^{i3,i4}_{\mathrm{CF+SO}}\delta^{i1,i2} + \mathcal{H}^{i1,i2}_{\mathrm{CF+SO}}\delta^{i3,i4} - \mathcal{H}^{i3,i2}_{\mathrm{CF+SO}}\delta^{i1,i4}-\mathcal{H}^{i1,i4}_{\mathrm{CF+SO}}\delta^{i2,i3}
\end{equation}
\noindent  for any two Slater determinants $n=(i1,i3)$ and $m=(i2,i4)$:
\begin{equation}
\mathcal{G}^{n}_{jj}(1,2) = \frac{1}{\sqrt{2}}\big[\psi_{i1}(1)\psi_{i3}(2)-\psi_{i1}(2)\psi_{i3}(1)\big]  \qquad \mathcal{G}^{m}_{jj}(1,2) = \frac{1}{\sqrt{2}}\big[\psi_{i2}(1)\psi_{i4}(2)-\psi_{i2}(2)\psi_{i4}(1)\big], 
\end{equation}
\noindent where $i1$, $i2$, $i3$, and $i4$ run over all orbitals in the order $\{d_{z^{2}}^{\uparrow},d_{z^{2}}^{\downarrow},d_{xy}^{\uparrow},d_{xy}^{\downarrow},d_{x^{2}-y^{2}}^{\uparrow},d_{x^{2}-y^{2}}^{\downarrow}\}$, and $\delta$ is the Kronecker symbol. The calculated two-electron levels $E_{{j}M}$ are shown in Fig.~\ref{fig:2e}. In the case of a perfect cubic environment (when $\hat{\mathcal{H}}_{\mathrm{CF}}+\hat{\mathcal{H}}_{\mathrm{SO}}$ is neglected), the two-electron states are split into three groups: $^3\mathrm{T}_1$, degenerate $^1\mathrm{T}_2$ and $^1\mathrm{E}$, and $^1\mathrm{A}_1$ with the energy levels $(U-5J)$, $(U-3J)$, and $U$, respectively~\cite{Oles2005}. This property can be used to derive the averaged Kanamori parameters~\cite{kanamori} for the low-temperature GaV$_{4}$S$_{8}$, $U = 0.858$ eV and $J = 0.087$ eV, in agreement with the constrained random phase approximation. 
\par Finally, $\hat{\mathcal{P}}_{j}$ is a projector operator that enforces the Pauli principle and suppresses any hoppings into the subspace of occupied orbitals at neighbouring sites. In practice, $\hat{\mathcal{P}}_{j}$ projects $|jM\rangle$ onto the subspace spanned by Slater determinants in the form:
\begin{equation}
\hat{\mathcal{P}}_{j}|jM(1,2)\rangle = \frac{1}{\sqrt{2}}\big[\psi_{j}(1)\phi_{j}(2)-\psi_{j}(2)\phi_{j}(1)\big],
\end{equation}
\noindent where $\phi_{j}$ is the occupied orbital at site $j$, and $\psi_{j}$ is an unoccupied orbital at site $j$ reached by the occupied electron $\phi_{i}$ through $\hat{t}_{ij}$.
\par To construct $\tensor{\mathcal{T}}_{ij}$, we consider all possible spin projections along $\pm x$, $\pm y$, and $\pm z$ for the occupied states at sites $i$ and $j$,  and the spin model is derived as~\cite{solse1,solse2}:
\begin{equation}
\mathcal{H}^{\mathrm{S}}=\sum\limits_{\langle ij\rangle} \boldsymbol{e}_{i}\tensor{\mathscr{J}}_{ij}\boldsymbol{e}_{j},  \qquad\qquad \mathscr{J}_{ij}^{ab}=\frac{1}{2}\big(\mathcal{T}^{ab}_{ij}-\mathcal{T}^{\bar{a}b}_{ij} \big),
\end{equation} 
\noindent where $\boldsymbol{e}_{i}=\langle\phi_{i}|\hat{\boldsymbol{\sigma}}|\phi_{i}\rangle/|\langle\phi_{i}|\hat{\boldsymbol{\sigma}}|\phi_{i}\rangle|$, $a$ and $b$ run over $\{x, y, z\}$, and $\bar{a}$ stands for the opposite direction of $a$. The anisotropic and symmetric parts can be rewritten as $\boldsymbol{e}_{i}\tensor{\mathcal{J}}_{ij}^{A}\boldsymbol{e}_{j}=\boldsymbol{D}_{ij}\cdot(\boldsymbol{e}_{i}\times\boldsymbol{e}_{j})$ and $\boldsymbol{e}_{i}\tensor{\mathcal{J}}_{ij}^{S}\boldsymbol{e}_{j}=-J_{ij}\boldsymbol{e}_{i}\cdot\boldsymbol{e}_{j}+\boldsymbol{e}_{i}\tensor{\Gamma}_{ij}\boldsymbol{e}_{j}\approx J_{ij}(e_{i}^{x}e_{j}^{x}+e_{i}^{y}e_{j}^{y})+(J_{ij}+\delta J_{ij})e_{i}^{z}e_{j}^{z}$, respectively. The resulting  parameters are summarized in Table~VII. The calculated values of $\delta J$ are 0.021 meV and 0.032 meV for the intraplane and interplane bonds, respectively, and can be neglected on account of their smallness.

\begin{figure*}
\begin{center}
\includegraphics[width=0.6\textwidth]{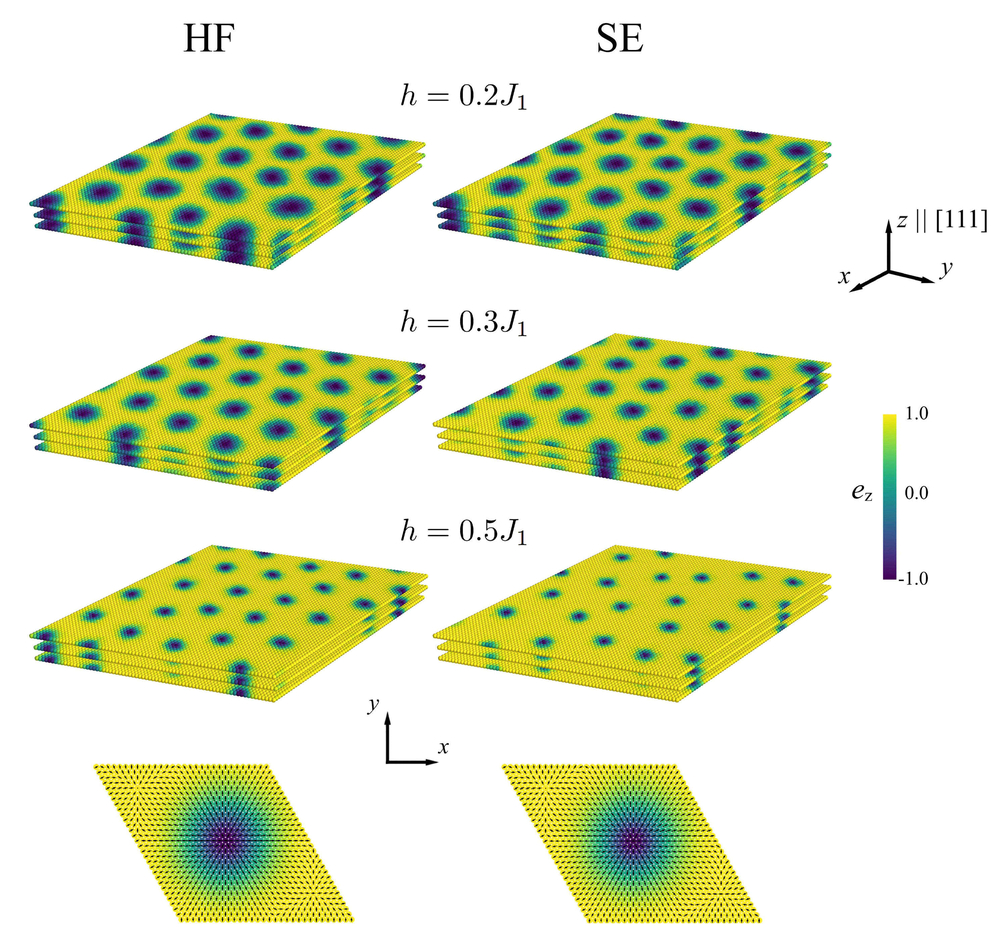}
\end{center}
\caption{Skyrmion lattices as obtained from Monte Carlo simulations at temperature $T=0.1J_{1}$ with an applied magnetic field $h\parallel[111]$ by using two sets of exchange parameters derived from Hartree-Fock calculations (HF) and superexchange theory (SE). }
\label{fig:mcres}
\end{figure*}
\subsection{Monte Carlo calculations}
\par Based on the calculated model parameters, we performed Monte Carlo simulations with an external magnetic field:
\begin{equation}
\mathcal{H} = \mathcal{H}^{\mathrm{S}} - h\sum_{i}e_{z}
\end{equation}
\noindent by using a heat-bath algorithm combined with overrelaxation~\cite{montecarlo}. In these calculations, supercells of $N=72\times72\times3$ spins with periodic boundary conditions were used, and a single run contained $(1.0-5.0)\times10^{6}$ Monte Carlo steps. For initial relaxation, the system was gradually cooled down from higher temperatures. As shown in Fig.~\ref{fig:mcres}, the resulting skyrmions are in agreement for two sets of the exchange parameters.

\section{Electric polarization}
\subsection{Berry-phase formalism}
\par Starting from the Hartree-Fock approximation, electric polarization can be calculated by using the Berry-phase formalism \cite{ber1, ber2}:
\begin{equation}
P_{a}=-\frac{1}{V}\frac{N_{a}}{N_{1}N_{2}N_{3}|\mathbf{G}_{a}|}\gamma_{a},
\end{equation}
\noindent where $V$ is the unit cell volume, and the Berry phase $\gamma_{a}$ is computed on a discrete mesh of $N_{1}\times N_{2}\times N_{3}$ $\boldsymbol{k}$-points in the first Brillouin zone with the reciprocal lattice vectors $\{\mathbf{G}_{a}\}$:
\begin{equation}
\boldsymbol{k}_{s1,s2,s3}=\frac{s_{1}}{N_{1}} \mathbf{G}_{1}+\frac{s_{2}}{N_{2}} \mathbf{G}_{2}+\frac{s_{3}}{N_{3}} \mathbf{G}_{3},
\end{equation}
\noindent and
\begin{equation}
\gamma_{1}=-\sum\limits_{s_{2}=0}^{N_{2}-1}
\sum\limits_{s_{3}=0}^{N_{3}-1}\mathrm{Im}\,\mathrm{ln}\prod\limits_{s_{1}=0}^{N_{1}-1}\mathrm{det}\,\hat{S}(\boldsymbol{k}_{s1,s2,s3},\boldsymbol{k}_{s1+1,s2,s3}),
\end{equation}
\noindent where $S_{nn'}(\boldsymbol{k},\boldsymbol{k'})=\langle\varphi_{n\boldsymbol{k}}|\varphi_{n'\boldsymbol{k'}}\rangle$ is an overlap matrix between occupied states at two neighbouring $\boldsymbol{k}$-points; $\gamma_{2}$ and $\gamma_{3}$ are obtained by cyclic permutation of $\{s_{1},s_{2},s_{3}\}$.  The on-site contribution to electric polarization can be calculated as:
\begin{equation}
\boldsymbol{P}^{\,on-site} = -\frac{e}{V} \,\mathrm{Tr}[\hat{n}\hat{\boldsymbol{r}}],
\end{equation}
\noindent where $\hat{\boldsymbol{r}}$ is the position operator defined in the Wannier function basis, and the trace is over site, orbital, and spin indices~\cite{ber3}.
\begin{figure*}
\begin{center}
\includegraphics[width=0.90\textwidth]{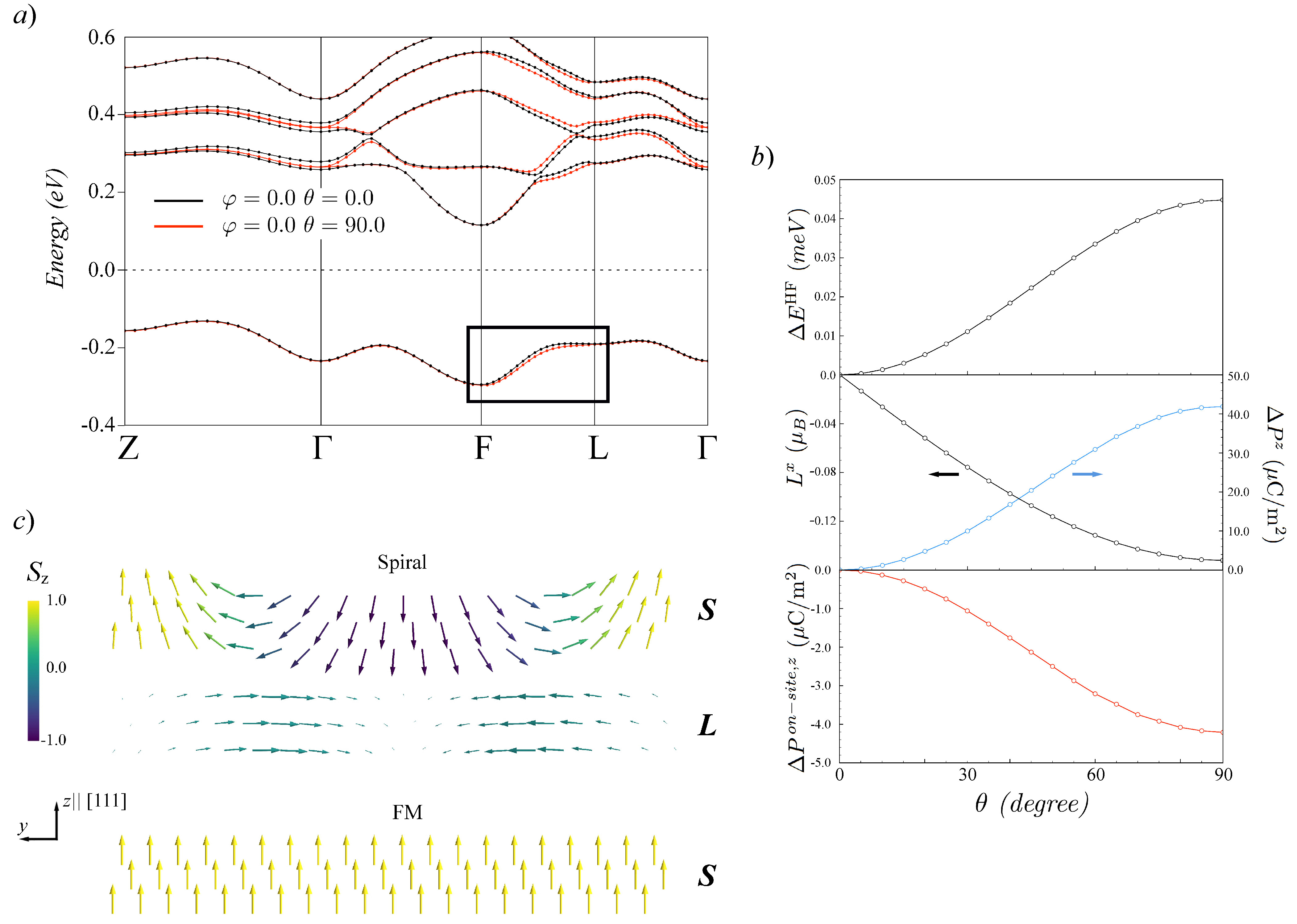}
\end{center}
\caption{$a$)~Band structures of the ferromagnetic configuration with $\theta=0^{\circ}$ and $\theta=90^{\circ}$ ($\phi=0^\circ$) as obtained in the Hartree-Fock approximation.~$b$) Angle dependence of the total energy, orbital angular momentum, electric polarization for the ferromagnetic configuration obtained in the Hartree-Fock approximation.~$c$) Spin and orbital (scaled for clarity) angular momenta of a spin spiral state obtained in the Hartree-Fock approximation with the propagation vector $\boldsymbol{q}=q(0,1,0)$ and a supercell $1\times20\times3$.}
\label{fig:hfrot}
\end{figure*}
\par The calculated values of electric polarization $P^{z}$  along $z\parallel[111]$ are 0.0107 and 0.0111 C/m$^{2}$ for the ferromagnetic solution with and without spin-orbit coupling, respectively. 
\par In the presence of spin-orbit coupling, we reveal a small change of electric polarization when the ferromagnetic moment is rotated away from the easy axis $z\parallel[111]$. Moreover, there is a strong anisotropy of orbital angular momentum in the direction perpendicular to the easy axis. It is worth noting that this fact is properly included in the following consideration of electric polarization in the framework of superexchange theory, where projections of the pseudospin state are constructed from the Kramers doublet $|\phi_{i}\rangle$ by diagonalising $\langle\phi_{i}|\hat{\boldsymbol{\sigma}}|\phi_{i}\rangle$. As seen from the band structure shown in Fig.~\ref{fig:hfrot}$a$ for two ferromagnetic solutions with different directions of the magnetization, the occupied band changes on the F--L path, which goes along the boundary of the Brillouin zone, that corresponds to the change of electron transfer between interplane nearest neighbours. Nevertheless, this change of electric polarization alongside with the on-site contribution is found to be small (Fig.~\ref{fig:hfrot}$b$) and can be neglected from further considerations for magnetic textures with a slowly varying magnetization profile.
\par Finally, we have performed Hartree-Fock calculations for the ferromagnetic and spin spiral states to calculate the change of electric polarization. The resulting magnetic textures are shown in Fig.~\ref{fig:hfrot}$c$, and the corresponding change of electric polarization between the ferromagnetic and spin spiral states is 10.4 and 12.1 $\mu$C/m$^{2}$ for supercells $1\times9\times3$ and $1\times12\times3$.

\subsection{Analytical expression without spin-orbit coupling}
\par The spin-dependent part of electric polarization can be described within perturbation theory \cite{depol,superpol}. To this end, let us consider the ferromagnetic state in the low-temperature phase, where a single $d_{z^{2}}^{\uparrow}$ state is occupied, and neglect spin-orbit coupling. Since all the calculated parameters of the corresponding low-energy model are close to those of the high-temperature phase, we can discard their orbital dependence, so that the on-site Hamiltonian is written as:
\begin{equation}
\hat{\mathcal{H}}_{0}=\hat{\mathcal{H}}_{\mathrm{CF}}+\hat{\mathcal{H}}_{U},
\end{equation}
\noindent where the crystal-field splitting between the $a_{1}$ and $e$ states and the Hubbard term are given by:
\begin{equation}
\hat{\mathcal{H}}_{\mathrm{CF}}=\sum_{i,m\in e,\sigma} \Delta\,\hat{c}_{im}^{\dagger\sigma}\hat{c}_{im}^{\sigma}
\end{equation}
\noindent and
\begin{equation}
\hat{\mathcal{H}}_{U}=\sum_{i,m}U\hat{n}_{im}^{\uparrow}\hat{n}_{im}^{\downarrow}+\sum_{i,\langle mm' \rangle\sigma}(U-2J)\,\hat{n}_{im}^{\sigma}\hat{n}_{im'}^{\bar{\sigma}}+(U-3J)\,\hat{n}_{im}^{\sigma}\hat{n}_{im'}^{\sigma},
\end{equation}
\noindent respectively. The corresponding level splitting at each site is readily obtained: $\varepsilon^{\alpha}_{i}=\{0;U;U-3J+\Delta;U-2J+\Delta;U-3J+\Delta;U-2J+\Delta\}$ for the states $\{d_{z^{2}}^{\uparrow},d_{z^{2}}^{\downarrow},d_{xy}^{\uparrow},d_{xy}^{\downarrow},d_{x^{2}-y^{2}}^{\uparrow},d_{x^{2}-y^{2}}^{\downarrow}\}$, respectively. As seen from Fig.~\ref{fig:dos}$b$, where the level splitting obtained from Hartree-Fock calculations is shown, the atomic limit is a good starting approximation since $t \ll U$. Next, we make use of a general definition of electric polarization expressed in terms of the Wannier functions~\cite{ber2}:
\begin{equation}
\boldsymbol{P}=-\frac{e}{V}\sum\limits_{i\alpha}^{occ}\int\limits_{V}\boldsymbol{r}|w_{i}^{\alpha}(\boldsymbol{r})|^{2}d\boldsymbol{r},
\end{equation}
\noindent Here, we assume that all $w_{i}^{\alpha}(\boldsymbol{r})$ are well localized at sites $i$, and their tales at neighbouring sites $j$ are proportional to the corresponding transfer integrals from the occupied to unoccupied orbitals and can be obtained within perturbation theory to second order in $\hat{t}_{ij}$:
\begin{equation}
|w_{i}^{\alpha}(\boldsymbol{r})|^{2}=w_{ii}^{\alpha}\delta(\boldsymbol{r}-\boldsymbol{R}_{i})+\sum\limits_{j}w^{\alpha}_{ij}\delta(\boldsymbol{r}-\boldsymbol{R}_{j}),
\end{equation}
\noindent where $w^{\alpha}_{ii}$ is the on-site Wannier density, and $w^{\alpha}_{ij}$ is the corresponding Wannier weight transfer from site $i$ to site $j$:
\begin{equation}
w^{\alpha}_{ij}=\sum\limits_{\beta}^{unocc}\left|\frac{t_{ij}^{\alpha\beta}}{\varepsilon_{i}^{\alpha}-\varepsilon_{j}^{\beta}}\right|^{2}.
\label{eq:wanwei}
\end{equation}
\noindent Consequently, electric polarization $\boldsymbol{P}$ can be represented as the sum of bond contributions:
\begin{equation}
\boldsymbol{P}=\sum_{\langle ij\rangle}\boldsymbol{P}_{ij},
\label{eq:polbond1}
\end{equation}
\noindent where
\begin{equation}
\boldsymbol{P}_{ij}=-\frac{e}{V}\boldsymbol{\tau}_{ji}\sum_{\alpha}(w^{\alpha}_{ij}-w^{\alpha}_{ji}),
\label{eq:polbond}
\end{equation}
\noindent and $\boldsymbol{\tau}_{ji}=\boldsymbol{R}_{j}-\boldsymbol{R}_{i}$. In the case of the low-temperature GaV$_{4}$S$_{8}$ with one occupied orbital $d_{z^{2}}^{\uparrow}$, $w^{\alpha}_{ij}$ can be expressed as:
\begin{equation}
w_{ij}=|\xi_{ij}^{\uparrow\uparrow}|^{2}\frac{(t_{ij}^{12})^{2}+(t_{ij}^{13})^{2}}{(U-3J+\Delta)^{2}}+|\xi_{ij}^{\uparrow\downarrow}|^{2}\frac{(t_{ij}^{12})^{2}+(t_{ij}^{13})^{2}}{(U-2J+\Delta)^{2}}+|\xi_{ij}^{\uparrow\downarrow}|^{2}\frac{(t_{ij}^{11})^{2}}{U^{2}}
\end{equation}
\noindent with
\begin{equation}
\begin{aligned}
|\xi_{ij}^{\uparrow\uparrow}|^{2}&=|\xi_{ij}^{\downarrow\downarrow}|^{2}=\frac{1}{2}(1+\boldsymbol{e}_{i}\cdot\boldsymbol{e}_{j}),\\
|\xi_{ij}^{\downarrow\uparrow}|^{2}&=|\xi_{ij}^{\uparrow\downarrow}|^{2}=\frac{1}{2}(1-\boldsymbol{e}_{i}\cdot\boldsymbol{e}_{j}),
\end{aligned}
\end{equation}
\noindent and the spin-dependent part of electric polarization is written as:
\begin{equation}
\boldsymbol{P}_{ij}=-\frac{e\boldsymbol{\tau}_{ji}}{2V}\left(\frac{1}{(U-3J+\Delta)^{2}}-\frac{1}{(U-2J+\Delta)^{2}}\right)\big((t_{ij}^{12})^{2}+(t_{ij}^{13})^{2}-(t_{ji}^{12})^{2}-(t_{ji}^{13})^{2}\big) \boldsymbol{e}_{i}\cdot\boldsymbol{e}_{j}.
\end{equation}
\noindent As follows, without spin-orbit coupling $\boldsymbol{P}_{ij}$ is entirely isotropic. This analysis leads to some important conclusions on the microscopic origin of electric polarization in GaV$_{4}$S$_{8}$. On the one hand, $\boldsymbol{P}_{ij}$ is given by the asymmetric electron transfer between nearest neighbors as a result of inversion symmetry breaking. On the other hand, $\boldsymbol{P}_{ij}$ requires a nonzero value of $J$. The corresponding values of $\boldsymbol{P}_{ij}$ are obtained by using the model parameters given in Tables IV and V: for the intraplane bonds $j=1$ -- $6$ (Fig.~\ref{fig:spin}$a$) $P_{01}=P_{02}=$-4.4 $\mu$C/m$^{2}$, $P_{04}=P_{05}=$4.4 $\mu$C/m$^{2}$, and $P_{03}=P_{06}=$0, for the interplane bonds $j=1'$ -- $6'$ (Fig.~\ref{fig:spin}$a$) we have $P_{0j}=(-1)^{j}P_{\perp}$, where $P_{\perp}=463.5$ $\mu$C/m$^{2}$. Thus, when summing over all nearest neighbors, the intraplane contributions cancel out, while the interplane contributions enter with opposite signs for the lower and upper planes and give the resulting spin-dependent polarization.

\begin{table*}[t]
\caption{Electric polarization tensors (in $\mu$C/m$^{2}$) calculated within superexchange theory for the $a_{1}\oplus e$ model of the low-temperature GaV$_{4}$S$_{8}$. The corresponding lattice translations are $\boldsymbol{a}_{1}=a(\sqrt{3}/2,-1/2,1.4251)$, $\boldsymbol{a}_{2}=a(0,1,1.4251)$, and $\boldsymbol{a}_{3}=a(-\sqrt{3}/2,-1/2,1.4251)$, where $a=3.9255$ \AA.}
\begin{center}
\begin{tabular}{ccc}
\hline
\hline
&& \\
\multicolumn{3}{c}{Intraplane} \\
&& \\
$
\tensor{\mathscr{P}}_{\bar{1}01}=\left(
\begin{array}{rrr}
       0.0 &  9.22 & -0.60 \\
       9.22 &  0.0 & 30.13 \\
      -0.60 &  -30.13 & 0.0 \\
\end{array} \right)
$ &
$
\tensor{\mathscr{P}}_{10\bar{1}}=\left(
\begin{array}{rrr}
       0.0 &  -9.22 &   0.60 \\
      -9.22 &   0.0 &   30.13 \\
       0.60 &  -30.13 &   0.0 \\
\end{array} \right)
$ &
$
\tensor{\mathscr{P}}_{\bar{1}10}=\left(
\begin{array}{rrr}
      -7.63 &    4.36 &  -26.20 \\
       4.83 &    8.31 &  -14.68 \\
      25.58 &   15.41 &    0.35 \\
\end{array} \right)
$  \\
&&\\
$
\tensor{\mathscr{P}}_{01\bar{1}}=\left(
\begin{array}{rrr}
      -7.63 & -4.36 &   26.20 \\
      -4.83 &  8.31 & -14.68 \\
     -25.58 & 15.41 &   0.35 \\
\end{array} \right)
$ &
$
\tensor{\mathscr{P}}_{1\bar{1}0}=\left(
\begin{array}{rrr}
       7.63 &   -4.83 &  -25.58 \\
      -4.36 &   -8.31 &  -15.41 \\
      26.20 &   14.68 &   -0.35 \\
\end{array} \right)
$ &
$
\tensor{\mathscr{P}}_{0\bar{1}1}=\left(
\begin{array}{rrr}
       7.63 &    4.83 &   25.58 \\
       4.36 &   -8.31 &  -15.41 \\
     -26.20 &   14.68 &  -0.35 \\
\end{array} \right)
$  \\
&&\\
\multicolumn{3}{c}{Interplane} \\
&&\\
$
\tensor{\mathscr{P}}_{00\bar{1}}=\left(
\begin{array}{rrr}
    -364.48 &   -6.91 &  -36.94 \\
      -6.22 & -356.96 &  -21.27 \\
      34.86 &   20.19 & -360.89 \\
\end{array} \right)
$ &
$
\tensor{\mathscr{P}}_{\bar{1}00}=\left(
\begin{array}{rrr}
    -364.48 &   6.91 &   36.94 \\
       6.22 & -356.96 &  -21.27 \\
     -34.86 &   20.19 & -360.89 \\
\end{array} \right)
$ &
$
\tensor{\mathscr{P}}_{001}=\left(
\begin{array}{rrr}
     364.48 &   6.22 & -34.86 \\
       6.91 & 356.96 &  -20.19 \\
      36.94 &  21.27 & 360.89 \\
\end{array} \right)
$  \\
&&\\
$
\tensor{\mathscr{P}}_{100}=\left(
\begin{array}{rrr}
     364.48 &  -6.21 & 34.86 \\
      -6.91 &  356.96 &  -20.19 \\
     -36.94 &   21.27 &  360.89 \\
\end{array} \right)
$ &
$
\tensor{\mathscr{P}}_{010}=\left(
\begin{array}{rrr}
     353.99 & 0.0 &  0.0 \\
       0.0 &  369.16 &   39.88 \\
       0.0  & -42.87 &  361.84 \\
\end{array} \right)
$ &
$
\tensor{\mathscr{P}}_{0\bar{1}0}=\left(
\begin{array}{rrr}
    -353.99 &  0.0 & 0.0 \\
       0.0 & -369.16 &   42.87 \\
       0.0 &  -39.88 & -361.84 \\
\end{array} \right)
$  \\
&&\\
\hline
\hline
\end{tabular}
\end{center}
\end{table*}

\subsection{Superexchange theory}
\par Following the idea of superexchange theory for magnetic interactions discussed above, we can consider hopping processes of a single occupied electron to the excited two-electron levels at neighbouring sites to construct  the Wannier weight transfer in Eq.~(\ref{eq:wanwei}):
\begin{equation}
\tensor{w}_{ij} = \left\langle \mathcal{G}_{ij}\left| \hat{t}_{ij} \left( \sum_M \frac{{\hat{\mathcal{P}}}_{j}|jM
\rangle \langle jM| {\hat{\mathcal{P}}}_{j}}{E^{2}_{jM}} \right) \hat{t}_{ji} \right| \mathcal{G}_{ij} \right\rangle,
\end{equation}
\noindent and the spin-dependent part of electric polarization in Eq.~(\ref{eq:polbond1}) can be written as:
\begin{equation}
\boldsymbol{P}_{ij}=\frac{\boldsymbol{\tau}_{ji}}{|\boldsymbol{\tau}_{ji}|}\boldsymbol{e}_{i}\tensor{\mathscr{P}}_{ij}\boldsymbol{e}_{j},
\end{equation}
\noindent where $\tensor{\mathscr{P}}_{ij}=||\mathscr{P}_{ij}^{ab}||$:
\begin{equation}
\mathscr{P}^{ab}_{ij}=-\frac{e|\boldsymbol{\tau}_{ji}|}{2V}\left(w_{ij}^{ab}-w_{ij}^{\bar{a}b} - w_{ji}^{ba} + w_{ji}^{b\bar{a}}\right),
\end{equation}
\noindent contains the isotropic $P_{ij}$, antisymmetric $\boldsymbol{\mathcal{P}}_{ij}$, and anisotropic $\tensor{\Pi}_{ij}$ components. The resulting tensors $\tensor{\mathscr{P}}_{ij}$ calculated from the multiplet structure with $\hat{\mathcal{H}}_{\mathrm{CF}}+\hat{\mathcal{H}}_{\mathrm{SO}}+\hat{\mathcal{H}}_{U}$ (Fig.~\ref{fig:2e}) are given in Table~VII. When taking into account the effects of crystal field and Coulomb interactions while neglecting spin-orbit coupling (the multiplet structure with $\hat{\mathcal{H}}_{\mathrm{CF}}+\hat{\mathcal{H}}_{U}$ in Fig.~\ref{fig:2e}), $\tensor{\mathscr{P}}_{ij}$ is diagonal and is in good agreement with the analytical one-electron approach: for the intraplane bonds $j=1$ -- $6$ (Fig.~\ref{fig:spin}$a$) $P_{0j}\approx0$, and for the interplane bonds $j=1'$ -- $6'$ (Fig.~\ref{fig:spin}$a$) $P_{0j}=(-1)^{j}P_{\perp}$, where $P_{\perp}=381.1$ $\mu$C/m$^{2}$.
\begin{figure*}
\begin{center}
\includegraphics[width=0.98\textwidth]{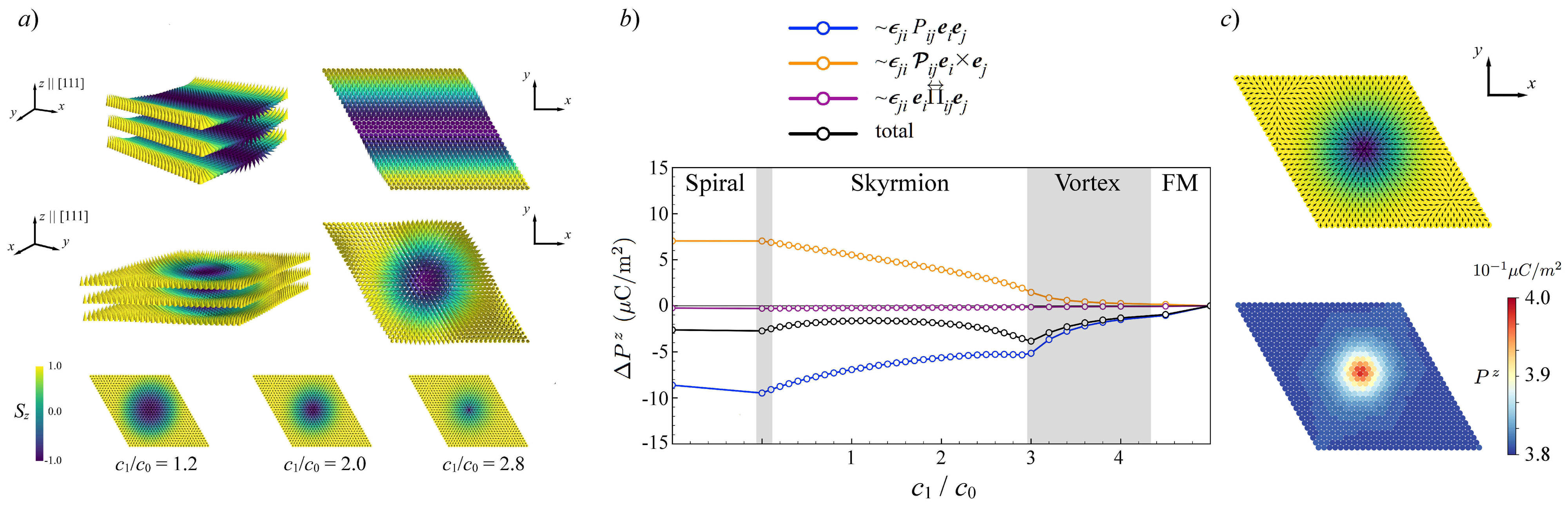}
\end{center}
\caption{$a$) Spiral and skyrmionic textures in the continuum limit. $b$) Change of electric polarization as a function of the skyrmionic radius calculated for the supercell $30\times30\times3$ with respect to the ferromagnetic state. $c$) Spatial profile of electric polarization in the skyrmionic phase.}
\label{fig:smpol}
\end{figure*}
\par To supplement the results of Monte Carlo calculations presented in the main text, we use the continuum limit  for magnetic textures, which is valid owing to their large spatial modulation. A skyrmionic texture can be defined as:
\begin{equation}
\boldsymbol{S}_{\boldsymbol{R}}=\frac{1}{2}\left(c_{0}\boldsymbol{S}_{0}+c_{1}\sum\limits_{n=1}^{3}\mathrm{Re}\,\left\{\boldsymbol{S}_{n}e^{i\boldsymbol{q}_{n}\cdot\boldsymbol{R}}\right\} \right),
\end{equation} 
\noindent where $\boldsymbol{S}_{n} = \boldsymbol{S}_{0}+i\boldsymbol{q}_{n}/q$ with $\boldsymbol{S}_{0}=(0,0,1)$, the propagation vectors of a skyrmionic texture are $\boldsymbol{q}_{1}=q(0,1,0)$, $\boldsymbol{q}_{2}=q(\sqrt{3}/2,-1/2,0)$, and  $\boldsymbol{q}_{3}=q(-\sqrt{3}/2,-1/2,0)$ with $q=2\pi/R_{\mathrm{SK}}$, $R_{\mathrm{SK}}$ is the skyrmion radius. For the spin spiral state, we have:
\begin{equation}
\boldsymbol{S}_{\boldsymbol{R}}=\frac{c_{1}}{2}\mathrm{Re}\,\left\{\boldsymbol{S}_{n}e^{i\boldsymbol{q}_{n}\cdot\boldsymbol{R}}\right\}.
\end{equation} 
\noindent Generally, the coefficients $c_{0}$ and $c_{1}$ depend on temperature and applied magnetic fields. In accordance with 
 Monte Carlo calculations, we consider supercells with a minimal periodicity along $z\parallel [111]$, as shown in Fig.~\ref{fig:smpol}$a$. The resulting change of spin-driven polarization as a function of $c_{1}/c_{0}$ is shown in Fig.~\ref{fig:smpol}$b$ and demonstrates a strong competition of the isotropic and antisymmetric contributions. The corresponding spatial profile of electric polarization in the skyrmionic phase is shown in Fig.~\ref{fig:smpol}$c$ and indicates a strong modulation of spin-driven polarization in the vicinity to the skyrmionic core, in agreement with Ref.~\onlinecite{pol2}.
